\documentclass[preprint,superscriptaddress,preprintnumbers,frontmatterverbose,amsmath,amssymb,aps,prd,floatfix,notitlepage,nofootinbib]{revtex4-2}

\usepackage{graphicx}
\usepackage{dcolumn}
\usepackage{bm}
\usepackage{hyperref}
\usepackage{orcidlink}
\usepackage{multirow}
\usepackage{float}
\allowdisplaybreaks

\def \dofrhoK{3}
\def \chirhoK{3.6}
\def \pvalrhoK{0.31}
\def \chirhoKConstraint {26.1}
\def \pvalrhoKConstraint{$9.1\times10^{-6}$}
\def \sigmarhoKConstraint {4.4}
\def \dofVV {5}
\def \chiVV {39.2}
\def \pvalVV{$2.2\times10^{-7}$}
\def \sigmaVV {5.2}
\def \dofAll {20}
\def \chiAll {130}
\def \pvalAll {$<10^{-15}$}
\def \sigmaAll {$>7$}

\def \beq{\begin{equation}}
\def \eeq{\end{equation}}
\def \bea{\begin{eqnarray}}
\def \eea{\end{eqnarray}}
\def \bma{\begin{matrix}}
\def \ema{\end{matrix}}
\def \s{\sqrt{2}}
\def \({\left(}
\def \){\right)}
\def \[{\left[}
\def \]{\right]}
\def \nn{\nonumber}
\def \nl{\nn\\}

\def \Z2{\mathbb{Z}_2}

\def \hf{\frac{1}{2}}

\def \de{\delta}

\def \la{\lambda}

\def\roughly#1{\mathrel{\raise.3ex\hbox
{$#1$\kern-.75em\lower1ex\hbox{$\sim$}}}}

\def\bs{B_s^0}

\def\order{\lower 1.8ex \hbox{\LARGE\~{}}}
\def\btopik{B \to \pi K}

\def \nl{\nonumber\\}
\def \s{\sqrt{2}}

\def\btopik{B \to \pi K}

\begin{document}

\title{\boldmath Anomalies in hadronic $B \to VV$ decays}

\author{Bhubanjyoti Bhattacharya\,\orcidlink{0000-0003-2238-321X}}
\email{bbhattach@ltu.edu}
\affiliation{Department of Natural Sciences, \\ Lawrence Technological University, Southfield, Michigan 48075, USA}

\author{Marianne Bouchard\,\orcidlink{0009-0005-2885-7473}}
\email{marianne.bouchard.5@umontreal.ca}
\affiliation{Physique des Particules, Universit\'e de Montr\'eal, 1375 Avenue Th\'er\`ese-Lavoie-Roux, Montr\'eal, Quebec, Canada  H2V 0B3}

\author{Alexandre Jean\,\orcidlink{0009-0005-2354-3299}}
\email{alexandre.jean@durham.ac.uk}
\affiliation{Institute for Particle Physics Phenomenology, Department of Physics, Durham University, Durham DH1 3LE, United Kingdom}

\author{David London\,\orcidlink{0000-0002-4407-5624}}
\email{london@lps.umontreal.ca}
\affiliation{Physique des Particules, Universit\'e de Montr\'eal, 1375 Avenue Th\'er\`ese-Lavoie-Roux, Montr\'eal, Quebec, Canada  H2V 0B3}

\author{Ipsita Ray\,\orcidlink{0000-0002-5506-754X}}
\email{ipsitaray02@gmail.com}
\affiliation{Physique des Particules, Universit\'e de Montr\'eal, 1375 Avenue Th\'er\`ese-Lavoie-Roux, Montr\'eal, Quebec, Canada  H2V 0B3}

\date{\today}

\begin{abstract}

Recently, a fit of charmless $B\to PP$ decays ($B \in \{B^0, B^+, \bs\}$, $P \in \{ \pi, K, \eta, \eta' \}$) to the latest data was performed under the assumption of flavor SU(3) symmetry [SU(3)$_F$]. It was found that there is a $4.1\sigma$ disagreement with the SU(3)$_F$ limit of the Standard Model [$\rm SM_{SU(3)_F}$]. In this paper, we extend this analysis to charmless $B \to VV$ decays ($V \in \{\rho, K^*, \phi, \omega\}$). The fit examining $B \to \rho K^*$ decays, assuming only isospin symmetry, is found to be acceptable. When we fit to $B \to VV$ decays with $V \in \{ \rho, K^* \}$ within SU(3)$_F$, we find a \sigmaVV$\sigma$ discrepancy with SM$_{\rm{SU(3)}_F}$. Finally, when $B \to VV$ decays with $V \in \{ \rho, K^*, \phi, \omega \}$ are considered, the discrepancy grows to \sigmaAll$\sigma$. The theoretical input in this analysis is modest, so our results are quite rigorous, group theoretically, and hold almost exactly in the SU(3)$_F$ limit. Although it seems unlikely that the introduction of $\sim 30$\% SU(3)$_F$-breaking effects can account for this discrepancy, this must be verified.

\end{abstract}

\preprint{UdeM-GPP-TH-26-310, IPPP/26/40}

\maketitle
\clearpage

\section{Introduction}

Although the Standard Model (SM) of particle physics has been enormously successful in describing almost all experimental results to date, it is not complete. This is because it cannot account for several important observations: neutrino masses, dark matter, the baryon asymmetry of the Universe, etc. There must be physics beyond the SM.

There are two ways to search for this new physics (NP). First, one can try to directly produce new particles at high-energy colliders. Unfortunately, despite years of direct searches, no new particles have been observed at the LHC. Because of this, our only alternative is to use indirect searches, in which the (higher-order) effects of the NP manifest themselves in measurements of low-energy observables that disagree with the predictions of the SM.

Indirect hints of NP have been seen in hadronic $B$ decays. In Refs~\cite{Berthiaume:2023kmp, Bhattacharya:2025wcq}, $B\to PP$ decays were examined under the assumption of flavor SU(3) symmetry [SU(3)$_F$]. Here $B \in \{B^0, B^+, \bs\}$ and the pseudoscalar $P \in \{ \pi, K, \eta, \eta' \}$.  Fits to the latest data were performed, and it was found that, although the individual fits to $\Delta S=0$ or $\Delta S=1$ decays are good, the combined fit is very poor: there is a $4.1\sigma$ disagreement with the SU(3)$_F$ limit of the Standard Model [SM$_{\rm{SU(3)}_F}$]. This discrepancy can be removed by adding SU(3)$_F$-breaking effects, but 1000\% SU(3)$_F$ breaking is required, far larger than the $\sim 30\%$ SU(3)$_F$ breaking expected in the SM. 

$B \to VV$ decays, in which the vector meson $V \in \{\rho, K^*, \phi, \omega\}$, have been less studied in this context. However, there have been some recent analyses of these decay modes, using the SU(2) U-spin symmetry \cite{Choudhury:2026avj} and the perturbative QCD approach \cite{Chai:2026rhh}. For a number of years, there have been some puzzling results in these decays. Because the $V$ is a spin-1 particle, it has three polarizations, one longitudinal and two transverse. And because the $B$ has spin 0, the final $VV$ pair in $B \to VV$ comes in three helicity states: both $V$s are either longitudinal or in one of two transverse states. Naive factorization suggests that the longitudinal amplitude dominates and the transverse components suffer a helicity suppression, i.e., the longitudinal and transverse polarization fractions should respect $f_L \gg f_T$, where $f_L + f_T = 1$ (see the discussion at the beginning of Sec.~\ref{sec:BtoVV} for additional details). However, the polarization fractions have been measured in many $B \to VV$ decay modes, and this is not always found to be true. For example, in $B \to \rho\rho$ decays, $f_L > 0.7$ is observed \cite{HFLAV:2024ctg}, in line with expectations. But in $B^0_s \to K^{*0} {\bar K}^{*0}$, $f_L = 0.16$ is measured \cite{HFLAV:2024ctg}, completely at odds with the prediction that $f_T/f_L \ll 1$.

Of course, it may well be that naive factorization is untrustworthy in $B \to VV$ decays, i.e., perhaps there are large nonfactorizable effects. Even so, such effects should respect flavor symmetries. However, in $B^0 \to K^{*0} {\bar K}^{*0}$, it is found that $f_L = 0.61$ \cite{HFLAV:2024ctg}.
Since $B^0_s \to K^{*0} {\bar K}^{*0}$ and $B^0 \to K^{*0} {\bar K}^{*0}$ are related by  $s \leftrightarrow d$ exchange, i.e. by U-spin symmetry, the very different measured values of $f_L$ in these two decays indicate strong U-spin violation \cite{Alguero:2020xca, Biswas:2025drz}. (The measured values of $f_L$ and $f_T$ in these modes also differ significantly from their theoretical predictions \cite{Aleksan:2023rkc}.) These $B^0_{d,s} \to K^{*0} {\bar K}^{*0}$ decays are also related to $B \to \rho K^*$ decays by SU(3)$_F$ symmetry, but in these latter decays, the observed values of $f_L$ vary from $\sim 0.7$ ($B^+ \to \rho^0 K^{*+}$) to $\sim 0.2$ ($B^0 \to \rho^0 K^{*0}$) \cite{HFLAV:2024ctg}.

Having observed a significant discrepancy with the SM in $B \to PP$ decays under the assumption of flavor SU(3) symmetry, it is only natural to perform an SU(3)$_F$ analysis of $B \to VV$ decays, to see if a similar discrepancy is found. The very curious results in the measurements of $f_{L,T}$ in these decays make such a study even more compelling.
In this paper, we carry out an SU(3)$_F$ analysis of $B\to VV$ decays.

In the analysis of $B\to PP$ decays, the Wigner-Eckart theorem was used to express all the amplitudes in terms of SU(3)$_F$ reduced matrix elements (RMEs). (Equivalently, the amplitudes can be decomposed in terms of topological diagrams.) A fit was done to all the $B\to PP$ data using the (complex) RMEs as the theoretical unknowns in order to determine how well the SM$_{\rm{SU(3)}_F}$ can explain the data. As noted above, the fit results show a significant discrepancy.

The analysis of $B \to VV$ decays is done in essentially the same way, with one important difference: each helicity state of a given $B \to VV$ decays has its own RMEs. So there are three times as many unknown parameters in $B \to VV$ decays as there are in $B \to PP$ decays. Fortunately, there are also many more $B \to VV$ observables, so that a fit can be done.

In this paper, we perform three types of fits. First, we examine only $B \to \rho K^*$ decays using isospin symmetry. Second, we study all $B \to VV$ decays in which $V \in \{ \rho, K^* \}$ assuming SU(3)$_F$. Finally, we consider $B \to VV$ decays with $V \in \{ \rho, K^*, \phi, \omega \}$. As we will see the disagreement with the symmetry limit of the SM grows as more decays are included. While the fit to $B \to \rho K^*$ is acceptable, the fit to $B \to VV$ decays with $V \in \{ \rho, K^* \}$ shows a \sigmaVV$\sigma$ discrepancy with the SM$_{\rm{SU(3)}_F}$, and this becomes \sigmaAll$\sigma$ (!) when $V \in \{ \rho, K^*, \phi, \omega \}$. Clearly this merits more study, both theoretical and experimental.

We begin in Sec.~\ref{sec:BtoVV} with a discussion of the formalism of $B \to VV$ decays, namely the helicities and the angular analysis. This includes a description of the different types of observables that are available in these decays. In Sec.~\ref{sec:decomposition}, we work out the decomposition of the $B \to VV$ amplitudes in terms of both RMEs and diagrams. The fit results are given in Sec.~\ref{sec:fit}. We summarize our findings in Sec.~\ref{sec:conclusion}. Additional technical details associated with the analysis have been relegated to the appendixes. The decomposition of the decays in terms of RMEs and diagrams can be found in Appendixes \ref{app:RMEs} and \ref{app:diags}, respectively. The observables used for the analysis are presented in Appendix \ref{app:data}. The fit results are shown in Appendix \ref{app:fit_results} and a discussion on the significant contributions to the $\chi^2$ can be found in Appendix \ref{app:contrib_chi2}. In Appendix \ref{app:relations}, we count the number of degrees of freedom (d.o.f.) using relationships between observables.

\section{\boldmath $B \to V V$ decays: formalism} 
\label{sec:BtoVV}

In $B \to V V$ decays, the final-state vector mesons have spin 1, so the total spin of the final state is $S_{\rm tot} = 0, 1$ or 2. The decaying $B$ meson has spin 0; in order to conserve the total angular momentum, the final-state $V$s must have relative orbital angular momentum $L = 0, 1$ or 2. These correspond, respectively, to $s$-, $p$- and $d$-wave final states.

$B \to V V$ decays can therefore be described by three separate amplitudes. In the helicity basis, these are $A_+$, $A_0$, and $A_-$, which obey the following hierarchy in naive factorization \cite{Beneke:2006hg}
\beq
A_0 : A_{+} : A_{-} \sim  1 : \frac{\Lambda_{\text{QCD}}}{m_b} : \bigg(\frac{\Lambda_{\text{QCD}}}{m_b} \bigg)^2 ~.
\label{eq:hierarchy}
\eeq

\subsection{Transversity amplitudes}

In experimental analyses, the observables are defined in terms of three transversity amplitudes, one for each polarization of the final state (one longitudinal, two transverse). The transversity amplitudes are the components of the full amplitude in which the polarizations of the final-state vector mesons are either longitudinal ($A_0$), or transverse to their directions of motion and parallel ($A_\|$) or perpendicular ($A_\perp$) to one another. They can be expressed in terms of the helicity amplitudes as follows: 
\beq
A_\| = \frac{1}{\sqrt{2}} (A_+ + A_-) ~~,~~~~ A_{\perp} = \frac{1}{\sqrt{2}} (A_+ - A_-) ~.
\label{trans-hel}
\eeq
$A_0$ and $A_\|$ are linear combinations of $s$- and $d$-wave (parity even), while $A_\perp$ is pure $p$-wave (parity odd). 
From Eqs.~(\ref{eq:hierarchy}) and (\ref{trans-hel}), we expect $|A_0| \gg |A_\||, |A_\perp|$. This can be tested experimentally through the measurements of the longitudinal and transverse polarization fractions, defined below in Sec.~\ref{sec:observables}.

\subsection{Differential decay rate}

\begin{figure}[t]
\begin{center}
    \includegraphics[width=0.5\textwidth]{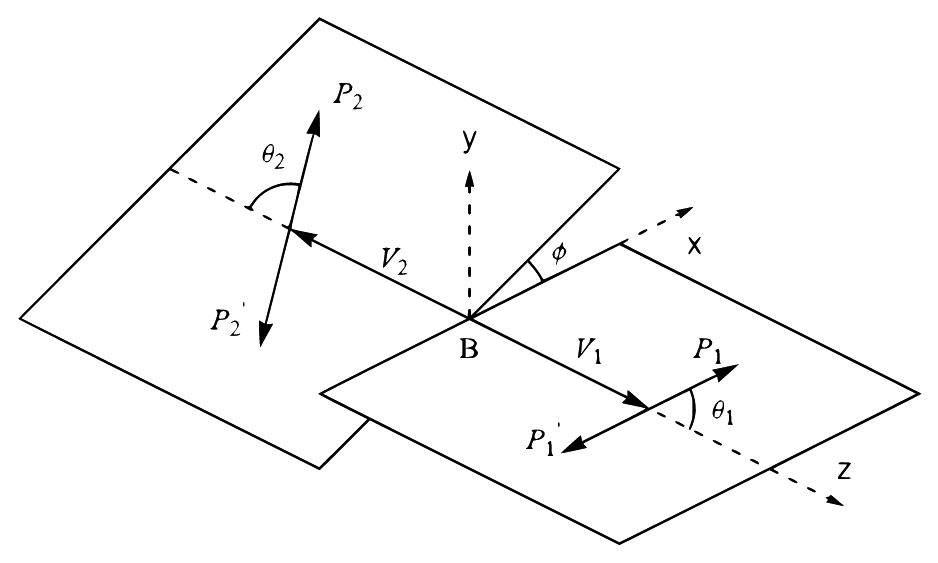} 
\end{center}
\caption{The decay $B \to V_1 (\to P_1 P'_1) ~ V_2 (\to P_2 P'_2)$.}
\label{BtoVVfig}
\end{figure}

Consider now the case where each final-state $V$ decays into two pseudoscalars. We have $B \to V_1 (\to P_1 P'_1) ~ V_2 (\to P_2 P'_2)$. This is shown in Fig.~\ref{BtoVVfig}. Here, three helicity angles are introduced: $\theta_1$ ($\theta_2$) is the angle between the directions of motion of the $P_1$ ($P_2$) in the $V_1$ ($V_2$) rest frame and the $V_1$ ($V_2$) in the $B$ rest frame, and $\phi$ is the angle between the normals to the planes defined by $P_1 P_1'$ and $P_2 P_2'$ in the $B$ rest frame. 

It was shown many years ago that one can separate the three helicities by performing an angular analysis of this decay \cite{Dighe:1995pd, Fleischer:1996aj, Dighe:1998vk, Chiang:1999qn, Bhattacharya:2013sga}. The angular distribution of this decay can be constructed as follows. Given that there are three helicities, when the amplitude is squared, there are six terms. If the decaying $B$ meson is neutral, there is time dependence due to $B^0_{(s)}$-${\bar B}^0_{(s)}$ mixing, but here we focus only on the angular distribution at time $t=0$. The differential decay rate is
\beq
\frac{d^3\Gamma}{d\cos\theta_1 d\cos\theta_2 d\phi} = \frac{9}{8 \pi} \sum^{6}_{i=1} K_i X_i(\theta_1,\theta_2,\phi) ~.
\label{Angdist}
\eeq
The $K_i$ and $X_i$ associated with the six terms of Eq.~(\ref{Angdist}) are given in Table \ref{tab:KX}. The measurement of the angular distribution allows the determination of the coefficients $K_i = \{ |A_0|^2$, $|A_\||^2$, $|A_\perp|^2$, ${\rm Re}[A_\|A^*_0]$, ${\rm Im}[A_\perp A^*_0]$, ${\rm Im}[A_\perp A^*_\|] \}$.

\begin{table}[h]
\centering
\begin{tabular}{|c|c|c|} \hline \hline
 $n$ & $K_i$ & $X_i$ \\ \hline
 1 & $|A_0|^2$     & $\cos^2\theta_1\cos^2\theta_2$     \\
 2 & $|A_\||^2$    & $\hf\sin^2\theta_1\sin^2\theta_2\cos^2\phi$ \\
 3 & $|A_\perp|^2$ & $\hf\sin^2\theta_1\sin^2\theta_2\sin^2\phi$ \\
 4 & ${\rm Re}[A_\|A^*_0]$& $\frac{1}{2\s}\sin2\theta_1\sin2\theta_2\cos\phi$ \\
 5 & ${\rm Im}[A_\perp A^*_0]$& $-\frac{1}{2\s}\sin2\theta_1\sin2\theta_2\sin\phi$ \\
 6 & ${\rm Im}[A_\perp A^*_\|]$& $-\hf\sin^2\theta_1\sin^2\theta_2\sin2\phi$ \\ 
 \hline\hline
\end{tabular}
\caption{Angular functions $X_i$ and their coefficients $K_i$ \cite{Bhattacharya:2013sga}.
\label{tab:KX}}
\end{table}

A similar angular distribution can be constructed for the $CP$-conjugate decay, but here the coefficients are functions of the $CP$-conjugate transversity amplitudes, ${\bar A}_\lambda$. These are obtained from the $A_\lambda$ by changing the signs of the weak phases. From the angular distribution of the $CP$-conjugate decay, one can determine the coefficients ${\bar K}_i = \{ |{\bar A}_0|^2$, $|{\bar A}_\||^2$, $|{\bar A}_\perp|^2$, ${\rm Re}[{\bar A}_\|{\bar A}^*_0]$, ${\rm Im}[{\bar A}_\perp {\bar A}^*_0]$, ${\rm Im}[{\bar A}_\perp {\bar A}^*_\|] \}$.

\subsection{Observables}
\label{sec:observables}

With the measurements of both angular distributions, one can obtain a variety of observables. First, there is the $CP$-averaged branching ratio, ${\cal B}_{CP}$:
\beq
{\cal B}_{CP} = F_{\rm PS} \, (|A|^2 + |{\bar A}|^2) ~, \label{eq:bcp}
\eeq
where $F_{PS} = \sqrt{m_B^2-(m_{V_1}+m_{V_2})^2} \, \sqrt{m_B^2-(m_{V_1}-m_{V_2})^2} \, S 
/ (32\pi m^3_B\Gamma_B)$ is the phase-space factor, and
\beq
|A|^2 = |A_0|^2 + |A_\||^2 + |A_{\perp}|^2 ~~,~~~~ 
|\bar{A}|^2 = |\bar{A}_0|^2 + |\bar{A}_\||^2 + |\bar{A}_{\perp}|^2 ~. 
\eeq
With these, the direct $CP$ asymmetry $\mathcal{A}_{CP}$ can be obtained:
\beq
\mathcal{A}_{CP} =\frac{|{\bar A}|^2-|A|^2}{|{\bar A}|^2+|A|^2} ~.
\label{directCPasym}
\eeq

With the measurements of $\{ |A_0|^2$, $|A_\||^2$, $|A_\perp|^2$, $|{\bar A}_0|^2$, $|{\bar A}_\||^2$, $|{\bar A}_\perp|^2 \}$, one can compute the polarization-dependent direct $CP$ asymmetry $\mathcal{A}_{CP}^\lambda$:
\beq
\mathcal{A}_{CP}^\lambda =\frac{|{\bar A}_\lambda|^2-|A_\lambda|^2}{|{\bar A}_\lambda|^2+|A_\lambda|^2} 
\label{poldirectCPasym}
\eeq
where $\lambda \in \{0,\|,\perp\}$ represents the polarization.
One can also measure the polarization fractions
\beq
f_{\lambda} = \frac{|A_{\lambda}|^2}{|A_0|^2 + |A_\||^2 + |A_{\perp}|^2} ~~,~~~~
{\bar f}_{\lambda} = \frac{|{\bar A}_{\lambda}|^2}{|{\bar A}_0|^2 + |{\bar A}_\||^2 + |{\bar A}_{\perp}|^2} ~. \label{eq:polf}
\eeq
With the $f_{\lambda}$, one can construct the longitudinal and transverse polarization fractions,
\beq
f_L = \frac{|A_0|^2}{|A_0|^2 + |A_\||^2 + |A_\perp|^2} ~~,~~~~
f_T = \frac{|A_\||^2 + |A_\perp|^2}{|A_0|^2 + |A_\||^2 + |A_\perp|^2} ~,
\eeq
and test whether the expectation that $|A_0| \gg |A_\||, |A_\perp|$, i.e., that $f_T/f_L \ll 1$, is borne out. (We will see that, for many decays, it is not.) $f_{\lambda}$ and ${\bar f}_{\lambda}$ can be combined to form the $CP$-averaged polarization fraction $\widetilde{f}_{\lambda}$ and the polarization fraction $CP$ asymmetry $\mathcal{A}_\lambda$:
\beq
\widetilde{f}_{\lambda} = \frac12 \, (f_{\lambda} + \bar{f}_{\lambda}) ~~,~~~~
\mathcal{A}_\lambda = \frac{\bar{f}_{\lambda}-f_{\lambda}}{\bar{f}_{\lambda}+f_{\lambda}} ~.
\label{polfracCPasym}
\eeq
Note that, in Ref.~\cite{LHCb:2025zvw}, the $CP$-averaged polarization fraction is defined differently:
\beq
f_{\la}^{\rm av} = \frac{|A_{\la}|^2+|\bar{A}_{\la}|^2}{\sum_{\lambda}(|A_{\lambda}|^2+|\bar{A}_{\lambda}|^2)} ~. \label{eq:polfCPavg}
\eeq
In Appendix \ref{app:relations}, we show that a polarization-dependent relationship exists between $f_\la^{av}$, ${\tilde f}_\la$, $\mathcal{A}_\lambda$, and $\mathcal{A}_{CP}$.

Another polarization-dependent measurement is the $CP$ violation parameter $|\la_i|$, defined as \cite{LHCb:2019jgw}
\beq
|\la_i| = \left|\frac{\bar A_i}{A_i}\right|~, \label{eq:ladef}
\eeq
where $i$ represents polarization.

The phases of the transversity amplitudes are neither weak nor strong, but rather a combination of the two. The measurements of the angular distributions allow for the extraction of various types of phase measurements: 
\bea
\delta_{\lambda} = \text{Arg}({A_{\lambda}})~, & ~~~~
\bar \delta_{\lambda} = \text{Arg}({\bar A_{\lambda}})~,~~~~
\widetilde\delta_{\lambda}^{} = \tfrac{1}{2}(\bar \delta_{\lambda} + \delta_{\lambda}) ~, & ~~~~
\delta_{\lambda}^{CP} = \tfrac{1}{2}(\bar \delta_{\lambda} - \delta_{\lambda})~.
\eea

Finally, mixing-induced indirect $CP$ violation has been measured in some of the $B \to VV$ decays, but  only for the longitudinal polarization component. The $CP$-violating observable is defined as
\begin{equation}
{\cal S}_{CP} = 2 \text{ Im}\bigg(\frac{q}{p} \frac{\bar{A}_0 A^*_{0}}{|\bar{A}_0|^2+|A_0|^2} \bigg) ~,
\end{equation}
where $q/p = {\rm exp}(-2 i \phi_M^q)$, in which $\phi_M^q$ is the weak phase of $B_q^0$-$\bar{B}_q^0$ ($q=d,s$) mixing.

Above, we have listed the observables that will be used in the fits. However, there are some other observables that merit a discussion.
Consider the direct $CP$ asymmetry $\mathcal{A}_{CP}$ of Eq.~(\ref{directCPasym}). $CP$-violating effects require the interference of (at least) two amplitudes with different weak phases. Assuming that two amplitudes contribute to the direct $CP$ asymmetry, we have $\mathcal{A}_{CP} \propto \sin \Delta\phi \sin \Delta\delta$, where $\Delta\phi$ and $\Delta\delta$ are the weak- and strong-phase differences between the two amplitudes. This shows that, in order to produce a nonzero direct $CP$ asymmetry, the two interfering amplitudes must also have different strong phases.

The observables that have coefficients ${\rm Im}[A_\perp A^*_0]$ and ${\rm Im}[A_\perp A^*_\|]$ represent T-odd ``triple products'' (TPs) of the form ${\vec p}_1 \cdot ({\vec p}_2 \times {\vec p}_3)$, where the ${\vec p}_i$ are the three-momenta of three of the final-state pseudoscalar mesons \cite{Datta:2003mj}. As was the case with $\mathcal{A}_{CP}$ [Eq.~(\ref{directCPasym})], $\mathcal{A}_{CP}^\lambda$ [Eq.~(\ref{poldirectCPasym})] and $\mathcal{A}_\lambda$ [Eq.~(\ref{polfracCPasym})], by combining the TPs in the two angular distributions, one can obtain a $CP$-violating observable. However, TPs are also P-odd, since they involve the interference of $A_\perp$ and $A_{0,\|}$. Because of this, the coefficients of the TPs in the $CP$-conjugate angular distribution are $-{\rm Im}[{\bar A}_\perp {\bar A}^*_0]$ and $-{\rm Im}[{\bar A}_\perp {\bar A}^*_\|]$. Thus, the ``true'' $CP$-violating TP is obtained by {\it adding} the TPs in the two distributions, while a ``fake'' $CP$-conserving TP is obtained by subtracting them. 

Assuming two interfering amplitudes, the TPs depend on $\Delta\phi$ and $\Delta\delta$ as follows:
\bea
{\hbox{True TPs}} &:& \Gamma + {\bar\Gamma} \propto \sin \Delta\phi \cos \Delta\delta ~, \nn\\
{\hbox{Fake TPs}} &:& \Gamma - {\bar\Gamma} \propto \cos \Delta\phi \sin \Delta\delta ~.
\eea
Note that a nonzero value of a $CP$-violating true TP does not require that the strong-phase difference $\Delta\delta$ be nonzero, in contrast to the direct $CP$ asymmetry $\mathcal{A}_{CP}$.

The true and fake TPs can be written as follows \cite{LHCb:2018hsm}:
\beq
\mathcal{A}^k_{\text{T-true}} = \frac12 (\mathcal{A}^k_T - \overline{\mathcal{A}}^k_T) ~~,~~~~ \mathcal{A}^k_{\text{T-fake}} = \frac12 (\mathcal{A}^k_T + \overline{\mathcal{A}}^k_T) ~,
\eeq
where $\mathcal{A}^1_T$ = $f_{\perp}f_0 \, \text{sin}(\delta_{\perp}-\delta_0)$ and $\mathcal{A}^2_T$ = $f_{\perp}f_\| \, \text{sin}(\delta_{\perp}-\delta_\|)$. The polarization fractions and phase differences can be extracted from measurements of the angular distributions, so the TPs are not independent observables. As a result, the measurements of the TPs themselves are not included in the fits. In Appendix \ref{app:relations}, we provide a detailed discussion of the number of observables used when performing our various fits.

\section{\boldmath $B \to V V$ decays: amplitude decomposition}
\label{sec:decomposition}

The aim of this paper is as follows. We consider all charmless $B \to VV$ decays, where $V \in \{\rho^\pm, \rho^0, K^{*\pm}, K^{*0}, {\bar K}^{*0}, \phi, \omega \}$. We assume a flavor symmetry group [isospin or flavor SU(3)], and we identify those decays whose amplitudes are related by this symmetry. We parametrize those amplitudes in terms of a certain number of unknown parameters and then perform a fit to the data to determine how well the SM can explain the experimental measurements in the limit of that symmetry.

The analysis can be performed in two ways. First, we can apply the Wigner-Eckart theorem associated with the respective symmetry group to express the amplitudes in terms of RMEs. We note that the weak Hamiltonian for charmless $B \to VV$ decay modes involves the quark-level transitions $\bar{b} \to \bar{u} u \bar{q}$ and ${\bar b} \to ({\bar q} d {\bar d}+{\bar q} s {\bar s})$ ($q = d, s$). It is given by \cite{Buchalla:1995vs}
\begin{equation} 
    H_W =  \frac{G_F}{\sqrt{2}}\sum_{q=d,s}\left(
\lambda_u^{(q)} \sum_{i=1}^2c_i Q_i^{(q)} - \lambda_t^{(q)} \sum_{i=3}^{10} c_i Q_i^{(q)}   \right) ~.
\label{eq:Hamiltonian}
\end{equation}
Here $\lambda_{q'}^{(q)} \equiv V_{q' b}^* V_{q' q}$ ($q' = u,c,t$, $q=d,s$), where the $V_{ij}$ are elements of the Cabibbo-Kobayashi-Maskawa (CKM) matrix. $Q_1^{(q)}$-$Q_2^{(q)}$, $Q_3^{(q)}$-$Q_6^{(q)}$, and $Q_7^{(q)}$-$Q_{10}^{(q)}$ represent tree, gluonic penguin, and electroweak penguin operators, respectively. The $c_i$s ($i = 1$-10) are the Wilson coefficients. In order to express the amplitudes in terms of RMEs, we must determine how each piece of the matrix element $\langle VV|H_W|B \rangle$ -- the initial state, the weak Hamiltonian, and the final state -- transforms under the symmetry. 

The second way of carrying out this analysis is to use topological diagrams \cite{Cheng:1986}. In
Refs.~\cite{Gronau:1994rj, Gronau:1995hn}, it was shown that the  amplitudes for charmless hadronic $B$ decays can be written in terms of  six diagrams: $T$ (tree), $C$ (color-suppressed
tree), $P$ (penguin), $A$ (annihilation), $E$ (exchange),
and $PA$ (penguin annihilation), of which $T$, $C$, $A$, and $E$ are proportional to $\lambda_u^{(q)}$. The diagrams $P$ and $PA$ each contain loops with internal up-type quarks, proportional to $\lambda_u^{(q)}$, $\lambda_{c}^{(q)}$, and $\lambda_t^{(q)}$. Using the unitarity of the CKM matrix, the $c$-quark pieces are eliminated to produce diagrams proportional to $\lambda_u^{(q)}$ ($P_{uc}$ and $PA_{uc}$) or $\lambda_t^{(q)}$ ($P_{tc}$ and $PA_{tc}$). In addition to these, there are also electroweak penguin (EWP) diagrams, all proportional to $\lambda_t^{(q)}$, as introduced in Ref.~\cite{Gronau:1998fn}. These are denoted as $P_{EW}$, $P^C_{EW}$, $P^{P_u}_{EW}$, $P^A_{EW}$, $P^E_{EW}$, and $P^{PA_u}_{EW}$. 

One of the advantages of diagrams over RMEs is that they represent different dynamical ways in which decays can be generated. As a result, one can estimate the relative sizes of various diagrams. This allows us to impose relations among diagrams, or even to neglect certain diagrams completely, reducing the number of unknown parameters. Here are some relations among diagrams that are often used:
\begin{itemize}

\item $E$, $A$, and $PA$ all require the $b$ quark to interact with the spectator quark. This implies that the diagrams are proportional to the decay constant $f_B$. This then implies that $E$/$A$/$PA$ are suppressed relative to other diagrams by $f_B/m_B \simeq 5\%$. Some analyses completely neglect $E$/$A$/$PA$, but it is perhaps more prudent to simply use the constraint $|E/T|, |A/T| \le 5\% $. $PA$ is expected to be even smaller, as it is suppressed due to the Okubo-Zweig-Iizuka rule.

\item The gluonic penguin diagram $P$ receives contributions from all three types of up-type quarks: 
$P = \lambda_{u}^{(q)} P_u + \lambda_{c}^{(q)} P_c + \lambda_{t}^{(q)} P_t = \lambda_{u}^{(q)} P_{uc} + \lambda_{t}^{(q)} P_{tc}$, where the second expression has been obtained using the unitarity of the CKM matrix ($\lambda_{u}^{(q)} + \lambda_{c}^{(q)} + \lambda_{t}^{(q)} = 0$), and $P_{uc} \equiv P_u - P_c$ and $P_{tc} \equiv P_t - P_c$. Because the $u$ and $c$ quarks are much closer to being degenerate than are the $t$ and $c$ quarks, it is expected that $|P_{uc}| < |P_{tc}|$. In Ref.~\cite{Buras:1994pb}, it is estimated that
\beq
0.2 \le |P_{uc}/P_{tc}| < 0.5 ~.
\label{Puc/Ptc}
\eeq
 
\item Naively, it is expected that $|C/T| = 1/3$, simply by counting colors. This ratio has been computed for $\btopik$ decays within QCD factorization, and $|C/T| \simeq 0.2$ is found at next-to-leading order (NLO) \cite{Beneke:2001ev}. The next-to-next-to-leading order analysis gives $0.13 \le |C/T| \le 0.43$, with a central value of $|C/T| = 0.23$, very near its NLO value \cite{Bell:2007tv, Bell:2009nk, Beneke:2009ek, Bell:2015koa}. For this reason, in some analyses, the constraint $|C/T| = 0.2$ is imposed.

\end{itemize}

In $B \to PP$ decays, under the assumption of SU(3)$_F$, there are five RMEs proportional to $\lambda_u^{(q)}$ and five proportional to $\lambda_t^{(q)}$. However, there are a total of six diagrams proportional to $\lambda_u^{(q)}$ and eight diagrams proportional to $\lambda_t^{(q)}$. The fact that there are more diagrams than RMEs implies that only certain linear combinations of diagrams appear in the amplitudes. 
These linear combinations can be designated as ``effective'' diagrams. This set of effective diagrams is equivalent to RMEs \cite{Gronau:1994rj}, i.e., each RME can be written as a linear combination of effective diagrams. As we will see, this same behavior occurs in $B \to VV$ decays -- in general, there are more diagrams than RMEs, so the effective diagrams must be identified.

Below, we present the decomposition of the $B \to VV$ amplitudes in terms of RMEs and diagrams for two different cases. First, we consider only $B \to \rho K^*$ decays, assuming isospin symmetry. Second, we consider all $B \to VV$ decays under the assumption of flavor SU(3) symmetry. 

\subsection{\boldmath $B \to \rho K^*$} 
\label{sec:isospin}

The ``$\btopik$ puzzle'' has been around for over 20 years. It involves the four decays  $B^+ \to \pi^+ K^{0}$, $B^+ \to \pi^0 K^{+}$, $B^0 \to \pi^- K^{+}$ and $B^0 \to \pi^0 K^{0}$, whose amplitudes obey a quadrilateral relation under isospin symmetry [SU(2)$_I$]:
\beq
A^{+0} + \s A^{0+} = A^{-+} + \s A^{00} ~,
\label{eq:isoquad}
\eeq
where the indices $i$ and $j$ in $A^{ij}$ indicate the charges of the final-state $\pi$ and $K$, respectively. It turns out that the measurements of the observables in these decays are not completely consistent with one another. This was first noted in Refs.~\cite{Buras:2003yc, Buras:2003dj, Buras:2004ub}. 

The four $B \to \rho K^*$ decay modes -- $B^+ \to \rho^+ K^{*0}$, $B^+ \to \rho^0 K^{*+}$, $B^0 \to \rho^- K^{*+}$ and $B^0 \to \rho^0 K^{*0}$ -- are just the excited versions of $\btopik$ decays. As such, their amplitudes are also related by isospin. More importantly, the expressions for the $B \to \rho K^*$ amplitudes in terms of RMEs/diagrams are identical to those of the corresponding $\btopik$ amplitudes. Recently, in Ref.~\cite{Bhattacharya:2025qye}, we reanalyzed the $\btopik$ puzzle. This involved expressing the four amplitudes in terms of RMEs/diagrams and performing a fit to the data. Below, we present these expressions for the four $\btopik$ amplitudes. 

In $\btopik$ decays, the initial state $B = (B^+, B^0)$ has isospin $\frac12$, and the final $\pi K$ states have isospin $\frac12$ or $\frac32$. These decays involve the transitions $\bar{b} \to \bar{u} u \bar{s}$ and $\bar{b} \to \bar{d} d \bar{s}$, so the weak Hamiltonian of Eq.~(\ref{eq:Hamiltonian}) has isospin 0 or 1 (with $I_3 = 0$). There are therefore three RMEs that describe these decays: $\langle \frac12 || H_W^0 || \frac12 \rangle$, $\langle \frac12 || H_W^1 || \frac12 \rangle$ and $\langle \frac32 || H_W^1 || \frac12 \rangle$. These RMEs can be split into pieces proportional to $\la_u^{(s)}$ and $\la_t^{(s)}$:
\bea
\langle \frac12 || H_W^0 || \frac12 \rangle = \la_u^{(s)}A_{0,1/2}^u -\la_t^{(s)}A_{0,1/2}^t~, \nn\\
\langle \frac12 || H_W^1 || \frac12 \rangle = \la_u^{(s)}A_{1,1/2}^u -\la_t^{(s)}A_{1,1/2}^t~, \\
\langle \frac32 || H_W^1 || \frac12 \rangle = \la_u^{(s)}A_{1,3/2}^u -\la_t^{(s)}A_{1,3/2}^t~, \nn
\eea
making a total of six RMEs.

In Ref.~\cite{Bhattacharya:2025qye}, it was shown that, if $c_7$ and $c_8$ are neglected (they are much smaller than $c_9$ and $c_{10}$), then for the two RMEs that involve $H_W^1$, (i) the $\la_u^{(s)}$ and $\la_t^{(s)}$ pieces are produced by the tree and EWP operators of the weak Hamiltonian, and (ii) these pieces are proportional to one another. To be specific, we have
\bea
A_{1,1/2}^t  = \frac{3}{2} \, \frac{c_9+c_{10}}{c_1+c_2} \, A_{1,1/2}^u ~~,~~~~
A_{1,3/2}^t  = \frac{3}{2} \, \frac{c_9+c_{10}}{c_1+c_2} \, A_{1,3/2}^u ~.
\label{eq:RMEEWPBpiK}
\eea
These are the EWP-tree relations at the level of RMEs, and they reduce the number of independent RMEs from six to four. It was argued in Ref.~\cite{Bhattacharya:2025qye} that, if $c_7$ and $c_8$ are not neglected, then these EWP-tree relations are changed by only $\sim 10\%$.

In terms of the SU(2)$_I$ RMEs, the $\btopik$ amplitudes are given by
\bea
A^{+0} &=& \lambda_{u}^{(s)} \[\sqrt{\frac{2}{3}} \(A_{0,1/2}^u-\frac{1}{\sqrt{3}} A_{1,1/2}^u \) +\frac{\sqrt{2}}{3} A_{1,3/2}^u\] \nn \\ 
&&\hspace{1truecm} -~\lambda_{t}^{(s)}\[\sqrt{\frac{2}{3}} \(A_{0,1/2}^t-\frac{1}{\sqrt{3}} A_{1,1/2}^t \) +\frac{\sqrt{2}}{3} A_{1,3/2}^t\] ~, \nn \\
\s\,A^{0+} &=& \lambda_{u}^{(s)} \[-\sqrt{\frac{2}{3}} \(A_{0,1/2}^u -\frac{1}{\sqrt{3}} A_{1,1/2}^u\) +\frac{2\sqrt{2}}{3} A_{1,3/2}^u \] \nn \\ 
&&\hspace{1truecm} -~\lambda_{t}^{(s)}\[-\sqrt{\frac{2}{3}} \(A_{0,1/2}^t -\frac{1}{\sqrt{3}} A_{1,1/2}^t\) +\frac{2\sqrt{2}}{3} A_{1,3/2}^t \] ~, \nn \\
A^{-+} &=& \lambda_{u}^{(s)} \[ -\sqrt{\frac{2}{3}} \(A_{0,1/2}^u+\frac{1}{\sqrt{3}} A_{1,1/2}^u \) +\frac{\sqrt{2}}{3} A_{1,3/2}^u\] \nn \\ 
&&\hspace{1truecm} -~\lambda_{t}^{(s)} \[ -\sqrt{\frac{2}{3}} \(A_{0,1/2}^t+\frac{1}{\sqrt{3}} A_{1,1/2}^t \) +\frac{\sqrt{2}}{3} A_{1,3/2}^t\]  ~, \nn \\
\sqrt{2} A^{00} &=& \lambda_{u}^{(s)} \[ \sqrt{\frac{2}{3}} \(A_{0,1/2}^u+\frac{1}{\sqrt{3}} A_{1,1/2}^u \) +\frac{2\sqrt{2}}{3} A_{1,3/2}^u\] \nn \\ 
&&\hspace{1truecm} -~\lambda_{t}^{(s)} \[ \sqrt{\frac{2}{3}} \(A_{0,1/2}^t+\frac{1}{\sqrt{3}} A_{1,1/2}^t \) +\frac{2\sqrt{2}}{3} A_{1,3/2}^t\]  ~,
\label{ampsSU(2)}
\eea
which satisfy Eq.~(\ref{eq:isoquad}). The amplitudes for the $CP$-conjugate processes can be obtained from the above by changing the sign of the weak phases.

In terms of diagrams, the $\btopik$ amplitudes are given by
\bea
A^{+0} &=& \lambda_{u}^{(s)} \widetilde{P_{uc}} 
+ \lambda_{t}^{(s)} \left[\widetilde{P_{tc}} - \frac{1}{3} \widetilde{P^C_{EW}} \right] ~, \nn\\
\s A^{0+} &=& \lambda_{u}^{(s)} \left[-\widetilde{T} - \widetilde{C} - \widetilde{P_{uc}} \right] 
 +~\lambda_{t}^{(s)} \left[-\widetilde{P_{tc}}-\widetilde{P_{EW}}-\frac{2}{3}\widetilde{P_{EW}^C}\right] ~, \nn\\
A^{-+} &=& \lambda_{u}^{(s)} \left[- \widetilde{T} - \widetilde{P_{uc}}\right]
+ \lambda_{t}^{(s)} \left[ - \widetilde{P_{tc}} - \frac{2}{3}\widetilde{P_{EW}^C} \right] ~, \nn\\
\s A^{00} &=& \lambda_{u}^{(s)} \left[-\widetilde{C} + \widetilde{P_{uc}} \right]  +~\lambda_{t}^{(s)} \left[\widetilde{P_{tc}} - \widetilde{P_{EW}} - \frac{1}{3} \widetilde{P_{EW}^C} \right] ~.
\label{eq:ampsred}
\eea
The six effective diagrams are defined as follows:
\bea
& \widetilde{T} \equiv T - A  ~,~~ \widetilde{C} \equiv C + A  ~,~~ \widetilde{P_{uc}} \equiv P_{uc} + A  ~,~  & \nn \\
& \widetilde{P_{EW}} \equiv P_{EW} + P_{EW}^E  ~,~~ \widetilde{P_{EW}^C} \equiv P_{EW}^C - P_{EW}^E  ~,~~  \widetilde{P_{tc}} \equiv P_{tc} - \frac{1}{3} (P_{EW}^{P_u}  - P_{EW}^E)  ~. &
\label{BpiKeffdiags}
\eea
At the level of diagrams, the EWP-tree relations are
\beq
\widetilde{P_{EW}} = -\frac{3}{2}\frac{c_9+c_{10}}{c_1+c_2} \, \widetilde{C} ~~,~~~~ 
\widetilde{P_{EW}^C} = -\frac{3}{2}\frac{c_9+c_{10}}{c_1+c_2} \, \widetilde{T} ~.
\label{eq:ETRSU2} 
\eeq
This reduces the number of independent effective diagrams from six to four.

To recap, the expressions for the four $\btopik$ amplitudes are given in terms of RMEs and diagrams in Eqs.~(\ref{ampsSU(2)}) and (\ref{eq:ampsred}), respectively. There are four independent RMEs/diagrams. These expressions also apply to each of the three helicities of the related $B \to \rho K^*$ amplitudes. For each polarization $\lambda$, the four amplitudes obey the isospin quadrilateral relation
\beq
A^{+0}_\lambda + \sqrt{2} A^{0+}_\lambda = \sqrt{2} A^{00}_\lambda + A^{-+}_\lambda ~.
\eeq

In Ref.~\cite{Gronau:2005kz}, an approximate sum rule involving the branching ratios and direct $CP$ asymmetries of the four $B \to \pi K$ decays was found under the assumption of SU(3)$_F$. In Ref.~\cite{Bhattacharya:2025qye}, it was shown that, if instead isospin symmetry is assumed, then this sum rule holds exactly. The experimental measurement by Belle II \cite{Belle-II:2023ksq} agrees with the prediction of the sum rule.

A similar sum rule applies to each polarization of the $B \to \rho K^*$ amplitudes. One constructs the rate difference as
\beq
\Delta^{ij}_{\lambda} ~=~ |{\overline A}^{ij}_{\lambda}|^2 - |A^{ij}_{\lambda}|^2\, \equiv \frac{{\cal A}_{CP}^{\lambda,ij} f^{av,ij}_{\lambda}{\cal B}_{CP}^{ij}}{F_{PS}^{ij}}~, 
\eeq
where ${\cal B}_{CP}^{ij}$, $A^{\lambda,ij}_{CP}$, $f^{av,ij}_{\lambda}$ and $F_{PS}^{ij}$ are, respectively, the $CP$-averaged branching ratio, the polarization-dependent direct $CP$ asymmetries, the $CP$-averaged polarization fractions, and the two-body phase-space factor for the decay $B \to \rho^i K^{*j}$.
 One can combine the $\Delta^{ij}_{\lambda}$s for the four decays and compute
\beq
\de^{\rho K^*}_{\lambda} ~=~ \Delta^{-+}_{\lambda} + \Delta^{+0}_{\lambda} - 2\,\Delta^{0+}_{\lambda} - 2\,\Delta^{00}_{\lambda}\,.
\label{sumrule}
\eeq
From the decomposition of the amplitudes in terms of RMEs [Eq.~(\ref{ampsSU(2)})] and diagrams [Eq.~(\ref{eq:ampsred})], one finds that $\de^{\rho K^*}_{\lambda} = 0$ is an exact consequence of applying the SU(2)$_I$ EWP-tree relations [Eqs.~(\ref{eq:RMEEWPBpiK}) and (\ref{eq:ETRSU2})] to each polarization of the $B \to \rho K^*$ decays.

We can combine the three polarizations: 
\beq
\Delta^{ij}= \Delta^{ij}_0+\Delta^{ij}_{||}+\Delta^{ij}_{\perp} ~\equiv~ \frac{{\cal A}_{CP}^{ij} {\cal B}_{CP}^{ij} }{F_{PS}^{ij}}~.
\eeq
 With this, one can compute the polarization-independent quantity
\beq
\de^{\rho K^*} ~=~ \Delta^{-+} + \Delta^{+0} - 2\,\Delta^{0+} - 2\,\Delta^{00} 
\label{totsumrule}
\eeq
for the four decays. Under isospin symmetry, one finds that $\de^{\rho K^*} = 0$ holds exactly. This sum rule can be tested experimentally: using the measurements in Appendix \ref{app:data}, we find that $\de^{\rho K^*} = (-4\pm6)\times10^{-4}$, which is consistent with zero.

We must stress that, in addition to the polarization-independent relation [Eq.~(\ref{totsumrule})], one can also experimentally test the relations separately for each transversity amplitude [Eq.~(\ref{sumrule})]. This will require a complete angular analysis of all four $B \to \rho K^*$ modes.

\subsection{\boldmath$B \to VV$} 
\label{sec:SU3}

We now consider all charmless $B \to VV$ decays, assuming flavor SU(3) symmetry. The SU(3)$_F$ octet contains the eight vector mesons $\{\rho^\pm, \rho^0, K^{*\pm}, K^{*0}, {\bar K}^{*0}, \omega_8 \}$, and $\omega_1$ is an SU(3)$_F$ singlet. The physical $\omega$ and $\phi$ mesons observed in experiments are admixtures of $\omega_8$ and $\omega_1$. The $B \to VV$ decay amplitudes can therefore be separated into three categories depending on the SU(3)$_F$ representation of the final-state vector mesons: $\boldsymbol{8\otimes8}$, $\boldsymbol{8\otimes1}$, and $\boldsymbol{1\otimes1}$. These are listed in Table \ref{tab:BVVdecays}.

\begin{table}
\begin{center}
\begin{tabular}{l c|c l}
    \multicolumn{4}{c}{$(\boldsymbol{8\otimes 8})_S$}\\
    \hline
    \multicolumn{2}{c|}{$\Delta S = 0$} &
    \multicolumn{2}{c}{$\Delta S = 1$}\\
    \hline\hline
    $B^+ \to K^{*+} \bar{K}^{*0}$ &&& $B^+\rightarrow \rho^+ K^{*0}$\\
    $B^+\rightarrow \rho^+\rho^0$ &&& $B^+\rightarrow \rho^0 K^{*+}$\\
    $B^+\rightarrow \omega_8 \, \rho^+$ &&& $B^+\rightarrow \omega_8 \, K^{*+}$\\
    \hline
    $B^0\rightarrow K^{*0}{\bar K^{*0}}$ &&& $\bs\rightarrow K^{*0}{\bar K^{*0}}$ \\
    $B^0\rightarrow \rho^+\rho^-$ &&&$\bs\rightarrow\rho^+\rho^-$ \\
    $B^0\rightarrow \rho^0\rho^0$&&& $\bs\rightarrow\rho^0\rho^0$\\
    $B^0\rightarrow K^{*+}K^{*-}$ &&& $\bs\rightarrow K^{*+}K^{*-}$\\
    $B^0\rightarrow \rho^0 \, \omega_8$ &&& $\bs\rightarrow\rho^0 \, \omega_8$\\
    $B^0\rightarrow \omega_8 \, \omega_8$&&& $\bs\rightarrow \omega_8 \, \omega_8$\\
    \hline
    $\bs\rightarrow\rho^+K^{*-}$&&&$B^0\rightarrow\rho^-K^{*+}$\\
    $\bs\rightarrow\rho^0{\bar K^{*0}}$&&&$B^0\rightarrow \rho^0 K^{*0}$\\
    $\bs\rightarrow\omega_8 \, {\bar K^{*0}}$&&&$B^0\rightarrow \omega_8 \, K^{*0}$\\
    \\
    \multicolumn{4}{c}{$\boldsymbol{8\otimes 1}$}\\
    \hline
    \multicolumn{2}{c|}{$\Delta S = 0$} &
    \multicolumn{2}{c}{$\Delta S = 1$}\\
    \hline\hline
    $B^+\rightarrow \omega_1 \, \rho^+$ &&& $B^+\rightarrow \omega_1 \, K^{*+}$\\
    \hline
    $B^0\rightarrow \rho^0 \, \omega_1$&&& $\bs\rightarrow \rho^0 \, \omega_1$\\
    $B^0\rightarrow \omega_8 \, \omega_1$&&& $\bs\rightarrow\omega_8 \, \omega_1$\\
    \hline
    $\bs\rightarrow\omega_1 \, {\bar K^{*0}}$&&&$B^0\rightarrow\omega_1 \, K^{*0}$\\
    \\
    \multicolumn{4}{c}{$\boldsymbol{1\otimes 1}$}\\
    \hline
    \multicolumn{2}{c|}{$\Delta S = 0$} &
    \multicolumn{2}{c}{$\Delta S = 1$}\\
    \hline\hline
$B^0\rightarrow \omega_1 \, \omega_1$&&& $\bs\rightarrow \omega_1 \, \omega_1$\\
\end{tabular}
\end{center}
    \caption{$\Delta S=0$ and $\Delta S=1$ $B \to VV$ decays for $V \in \{\rho, K^*, \omega_8, \omega_1\}$.}
    \label{tab:BVVdecays}
\end{table}

Within SU(3)$_F$, all members of the octet $\boldsymbol{8}$ are considered to be identical particles. In this case, the $\boldsymbol{8\otimes8}$ final state, which is the product of three pieces [SU(3)$_F$, spatial (orbital angular momentum), and spin], must be symmetric under the exchange of the two particles.
The decaying $B$ meson has spin 0, so the conservation of total angular momentum implies that,  when the spatial wave function of the final-state mesons is symmetric ($l = 0,2$), the spin wave function is also symmetric ($s = 0,2$). Similarly, when the spatial wave function is antisymmetric ($l = 1$), the spin wave function is also antisymmetric ($s = 1$). Thus, the product of final-state spatial and spin wave functions is always symmetric, which implies that the SU(3)$_F$ piece of the final state is also symmetric, $(\boldsymbol{8\otimes 8})_S$.

\subsubsection{RMEs} 

We begin with the $(\boldsymbol{8\otimes 8})_S$ final state. The quarks $(u,d,s)$ form a triplet ($\boldsymbol{3}$) under SU(3)$_F$, so that the initial state $B = (B^+, B^0, \bs)$ also is a $\boldsymbol{3}$. The weak Hamiltonian [Eq.~(\ref{eq:Hamiltonian})] transforms as a ${\bf 3^*_1}$, ${\bf 3^*_2}$, ${\bf 6}$, or ${\bf 15^*}$ of SU(3)$_F$ (for more details on these representations, we refer the reader to Refs.~\cite{Bhattacharya:2025wcq,Bhattacharya:2025qye}). For the final state, $(\boldsymbol{8 \otimes 8})_S = \boldsymbol{ 1 \oplus 8}_S \boldsymbol {\oplus 27}$. This yields five RMEs that can be split into pieces proportional to $\la_u^{(q)}$ and $\la_t^{(q)}$:
\bea
& A_1 = \langle {\bf 1} || {\bf 3^*} || {\bf 3} \rangle = \la_u^{(q)} A_1^u - \la_t^{(q)} A_1^t ~,~~
A_8 = \langle {\bf 8} || {\bf 3^*} || {\bf 3} \rangle = \la_u^{(q)} A_8^u - \la_t^{(q)} A_8^t ~, & \nn\\
& R_8 = \langle {\bf 8} || {\bf 6} || {\bf 3} \rangle = \la_u^{(q)} R_8^u -  \la_t^{(q)} R_8^t ~,~~ 
P_8 = \langle {\bf 8} || {\bf 15^*} || {\bf 3} \rangle = \la_u^{(q)} P_8^u - \la_t^{(q)} P_8^t ~, & \nn\\
& P_{27} = \langle {\bf 27} || {\bf 15^*} || {\bf 3} \rangle = \la_u^{(q)} P_{27}^u - \la_t^{(q)} P_{27}^t ~. &
\label{eq:8x8RMEs}
\eea
Here, the $A_{1,8}^u$ RMEs involve the ${\bf 3^*_1}$ piece of $H_W$, while the $A_{1,8}^t$ RMEs involve the ${\bf 3^*_2}$ piece. (Note that the notation used here for the $\langle {\bf 1} || {\bf 3^*} || {\bf 3} \rangle$ and $\langle {\bf 8} || {\bf 3^*} || {\bf 3} \rangle$ RMEs is different from that used in Refs~\cite{Berthiaume:2023kmp, Bhattacharya:2025wcq}: $A_{1(8)}^u$ and $A_{1(8)}^t$ in this paper correspond, respectively, to $A_{1(8)}$ and $B_{1(8)}$ of Refs.~\cite{Berthiaume:2023kmp, Bhattacharya:2025wcq}.)

The last three RMEs above have EWP-tree relations \cite{Bhattacharya:2025qye}:
\beq
R_8^t = \frac32 \, \frac{(c_9 - c_{10})}{(c_1 - c_2)} \, R_8^u ~~,~~~~
P_8^t = \frac32 \, \frac{(c_9 + c_{10})}{(c_1 + c_2)} \, P_8^u ~~,~~~~
P_{27}^t = \frac32 \, \frac{(c_9 + c_{10})}{(c_1 + c_2)} \, P_{27}^u ~.
\label{ETRsRMEs}
\eeq
This reduces the number of independent RMEs in $B \to PP$ decays from ten to seven. In the $B \to VV$ decays, the number is reduced from ten to seven for each polarization.

As was the case with $B \to \rho K^*$ and $\btopik$, each of the $(\boldsymbol{8\otimes8})_S$ $B \to VV$ decays in Table \ref{tab:BVVdecays} has a $B \to PP$ counterpart ($\rho \to \pi$, $K^* \to K$, $\omega_8 \to \eta_8$). These $B\to PP$ decays were analyzed in Ref.~\cite{Bhattacharya:2025wcq} under the assumption of SU(3)$_F$ symmetry. For these decays, the final state also transforms as $(\boldsymbol{8\otimes 8})_S$, so that the expressions for the $B \to PP$ amplitudes in terms of RMEs/diagrams also apply to each polarization of the related $B \to VV$ decay. The decomposition of the various $(\boldsymbol{8\otimes 8})_S$ $B \to VV$ amplitudes in terms of the SU(3)$_F$ RMEs defined in Eq.~(\ref{eq:8x8RMEs}) is shown in Appendix \ref{app:RMEs}.

The final states that transform as $\boldsymbol{8\otimes1}$ include one of $\rho$, $K^{*}$ or $\omega_8$, and $\omega_1$. There are three RMEs that can be split into pieces proportional to $\lambda_u^{(q)}$ and $\lambda_t^{(q)}$:
\bea
& C_8 = \langle {\bf 8} || {\bf 3^*} || {\bf 3} \rangle_{{\bf 8}\otimes{\bf 1}} = \lambda_u^{(q)} C_8^u - \lambda_t^{(q)} C_8^t~,  & \nn \\
& L_8 = \langle {\bf 8} || {\bf 6} || {\bf 3} \rangle_{{\bf 8}\otimes{\bf 1}} = \lambda_u^{(q)} L_8^u - \lambda_t^{(q)} L_8^t ~,~~ 
M_8 = \langle {\bf 8} || {\bf 15^*} || {\bf 3} \rangle_{{\bf 8}\otimes{\bf 1}} = \lambda_u^{(q)} M_8^u -\lambda_t^{(q)} M_8^t ~. &
\label{eq:8x1RMEs}
\eea
Here, the $C_8^u$ RME involves the ${\bf 3^*_1}$ piece of $H_W$, while the $C_8^t$ RME involves the ${\bf 3^*_2}$ piece.

The last two RMEs above have EWP-tree relations similar to those in Eq.~(\ref{ETRsRMEs}):
\beq
L_8^t = \frac32 \, \frac{(c_9 - c_{10})}{(c_1 - c_2)} \, L_8^u ~~,~~~~
M_8^t = \frac32 \, \frac{(c_9 + c_{10})}{(c_1 + c_2)} \, M_8^u ~.
\label{ETRsRMEs1}
\eeq
This reduces the number of independent RMEs in these decays from six to four. The amplitude decompositions for the $\boldsymbol{8\otimes1}$ $B \to VV$ decays in terms of the above RMEs are given in Appendix \ref{app:RMEs}.

Finally, the $\boldsymbol{1\otimes1}$ final states contain two $\omega_1$s. There is a single RME that can be split into pieces proportional to $\lambda_u^{(q)}$ and $\lambda_t^{(q)}$:
\bea
C_1 = \langle {\bf 1} || {\bf 3^*} || {\bf 3} \rangle_{{\bf 1}\otimes{\bf 1}} = \lambda_u^{(q)} C_1^u - \lambda_t^{(q)} C_1^t~,
\label{eq:1x1RMEs}
\eea
with no EWP-tree relations. There are therefore two independent RMEs. The amplitude decompositions for the $\boldsymbol{1\otimes1}$ $B \to VV$ decays can be found in Appendix \ref{app:RMEs}.

\subsubsection{Diagrams} 

The amplitudes can also be expressed in terms of diagrams. There are six diagrams proportional to $\lambda_u^{(q)}$: $T$, $C$, $P_{uc}$, $E$, $A$, $PA_{uc}$. There are eight diagrams proportional to $\lambda_t^{(q)}$: $P_{tc}$, $PA_{tc}$, $P_{EW}$, $P_{EW}^C$, $P_{EW}^A$, $P_{EW}^E$, $P_{EW}^{P_u}$, $P_{EW}^{PA_u}$. 

For the $(\boldsymbol{8\otimes 8})_S$ final state, Ref.~\cite{Bhattacharya:2025wcq} gives the expressions for the $B \to PP$ amplitudes in terms of diagrams. As usual, these expressions also apply to each polarization of the $B \to VV$ decays. They are given in Appendix \ref{app:diags}.

Since there are six diagrams proportional to $\lambda_u^{(q)}$, but only five RMEs [Eq.~(\ref{eq:8x8RMEs})]; only five linear combinations of the diagrams can appear in the amplitudes. We eliminate the $E$ diagram by defining five effective diagrams:
\bea
& \widetilde{T} = T + E ~~,~~~~
\widetilde{C} = C - E ~~,~~~~
\widetilde{P_{uc}} = P_{uc} - E ~, & \nn\\
& \widetilde{A} = A + E ~~,~~~~
\widetilde{PA}_{uc} = PA_{uc} + E ~. &
\label{effdiaglambdau}
\eea
Similarly, there are eight diagrams proportional to $\lambda_t^{(q)}$, but only five RMEs. Once again, only five linear combinations of the diagrams will appear in the amplitudes. To this end, we define five effective diagrams:
\bea
& \widetilde{P_{EW}} \equiv P_{EW} + P_{EW}^E ~~,~~~~
\widetilde{P_{EW}^C} \equiv P_{EW}^C - P_{EW}^E ~~,~~~~
\widetilde{P_{EW}^A} \equiv P_{EW}^{A} + P_{EW}^E ~, & \nn\\
& \widetilde{P_{tc}} \equiv P_{tc} - \frac{1}{3} (P_{EW}^{P_u}  - P_{EW}^E) ~~,~~~~
\widetilde{PA_{tc}} \equiv PA_{tc} - \frac{1}{3} ( P_{EW}^{PA_{u}} + P_{EW}^E) ~. &
\label{effdiaglambdat}
\eea
Only the effective diagrams of Eqs.~(\ref{effdiaglambdau}) and (\ref{effdiaglambdat}) appear in the $B \to VV$ amplitudes, and each $\lambda_u^{(q)}$ [$\lambda_t^{(q)}$] effective diagram can be written as a linear combination of the five $\lambda_u^{(q)}$ [$\lambda_t^{(q)}$] RMEs.

Finally, just as there were three EWP-tree relations involving the RMEs [Eq.~(\ref{ETRsRMEs})], there are three EWP-tree relations involving the effective diagrams:
\bea
\widetilde{P_{EW}} &=& -\frac{3}{4}\left[\frac{\Sigma_9}{\Sigma_1}(\tilde{T}+\tilde{C}+\tilde{A})+\frac{\Delta_9}{\Delta_1}(\tilde{T}-\tilde{C}-\tilde{A})\right] ~, \nn\\
\widetilde{P^C_{EW}} &=& -\frac{3}{4}\left[\frac{\Sigma_9}{\Sigma_1}(\tilde{T}+\tilde{C}-\tilde{A})-\frac{\Delta_9}{\Delta_1}(\tilde{T}-\tilde{C}-\tilde{A})\right] ~, \\
\widetilde{P^A_{EW}} &=& -\frac{3}{2}\frac{\Sigma_9}{\Sigma_1}\tilde{A} ~, \nn
\label{eq:symEWP}
\eea
where $\Sigma_1 = c_1+c_2$, $\Sigma_9= c_9+c_{10}$, $\Delta_1=c_1-c_2$, and $\Delta_9 = c_9-c_{10}$. 
With these EWP-tree relations, there are five independent effective diagrams in the $\lambda_u$ sector, $\widetilde{T}$, $\widetilde{C}$, $\widetilde{P_{uc}}$, $\widetilde{A}$ and $\widetilde{PA_{uc}}$, and two in the $\lambda_t$ sector, $\widetilde{P_{tc}}$ and $\widetilde{PA_{tc}}$.

The expressions for the $\boldsymbol{8\otimes1}$ $B \to PP$ amplitudes in terms of diagrams are computed in Ref.~\cite{Bhattacharya:2025wcq}. As usual, these expressions also apply to each polarization of the corresponding $B \to VV$ decays. They are given in Appendix \ref{app:diags}. We note that there are five diagrams in the $\lambda_u^{(q)}$ sector, but there are only three independent RMEs. The three effective diagrams that appear in the amplitudes are defined as
\bea
&&\widetilde{T} \equiv T+2A~,~~\widetilde{C} \equiv C+2E~,~~\widetilde{P_{uc}} \equiv P_{uc}-E ~.
\eea
Similarly, there are six diagrams proportional to $\lambda_t^{(q)}$ in the amplitudes, but they only appear in three linear combinations. The three independent effective diagrams are defined as
\bea
\widetilde{P_{EW}} \equiv P_{EW} + 2P_{EW}^A ~&,~ \widetilde{P_{EW}^C} \equiv P_{EW}^C + 2P_{EW}^E ~&,~
 \widetilde{P_{tc}} \equiv P_{tc} -\frac{1}{3} (P_{EW}^{P_u} - P_{EW}^E)~.
\eea

There are therefore two EWP-tree relations involving the effective diagrams:
\begin{equation}
\begin{split}
    \widetilde{P_{EW}} &= -\frac{3}{4}\left[\frac{\Sigma_9}{\Sigma_1}(\widetilde{T}+\widetilde{C}) + \frac{\Delta_9}{\Delta_1}(\widetilde{T}-\widetilde{C})\right] ~, \\
    \widetilde{P_{EW}^C} &=  -\frac{3}{4}\left[\frac{\Sigma_9}{\Sigma_1}(\widetilde{T}+\widetilde{C}) - \frac{\Delta_9}{\Delta_1}(\widetilde{T}-\widetilde{C})\right] ~. \\
\end{split}
\label{eq:EWPtree8x1}
\end{equation}
This leads to a total of three independent diagrams in the $\lambda_u$ sector, $\widetilde{T}$, $\widetilde{C}$ and $\widetilde{P_{uc}}$, and one in the $\lambda_t$ sector, $\widetilde{P_{tc}}$. [Note that, despite the fact that they have the same name, these diagrams are not the same as those in the $(\boldsymbol{8\otimes 8})_S$ amplitudes.]

Finally, using the expressions for the $\boldsymbol{1\otimes1}$ $B \to PP$ amplitudes in terms of diagrams given in Ref.~\cite{Bhattacharya:2025wcq},  the amplitude decompositions for the $\boldsymbol{1\otimes1}$ $B \to VV$ decays in terms of diagrams are shown in Appendix \ref{app:diags}. From these, we find that there is only one effective diagram proportional to $\lambda_u^{(q)}$, $\widetilde{C} \equiv C + P_{uc} + E + 6PA_{uc}$, and one proportional to $\lambda_t^{(q)}$, $\widetilde{P_{tc}} \equiv P_{tc} + 6PA_{tc} -\frac{1}{3}P_{EW}^C -\frac{1}{3}P_{EW}^E - \frac{1}{3}P_{EW}^{P_u} -2 P^{PA_u}_{EW}$. There is no EWP-tree relations in this case.

\section{Fit Results}
\label{sec:fit}

In this section we present the results of our fits to the data in $B \to VV$ decays. We describe three different analyses. First, we examine only $B \to \rho K^*$ decays, assuming isospin symmetry. Second, we combine all $B \to VV$ decays under the assumption of SU(3)$_F$, but restrict $V \in \{ \rho, K^* \}$.  Finally, we extend the SU(3)$_F$ analysis to include all $B \to VV$ modes, $V \in \{ \rho, K^*, \phi, \omega \}$. 

In the previous section, we expressed the $B \to VV$ amplitudes as functions of both RMEs and effective diagrams. As long as all diagrams are kept, these two are equivalent. We choose to perform the analyses using the effective diagrams as the unknown parameters to be determined from the fit. The advantage of using diagrams is that it allows us to easily add theoretical input by making assumptions about the relative sizes of the diagrams.

In addition to the diagrams (magnitudes and relative strong phases), the amplitudes depend on the CKM matrix elements involved in $\lambda_{u,t}^{(q)}$, and the weak phases $\beta$, $\gamma$ and $\beta_s$. The values of all of these quantities are fixed to their world averages \cite{ParticleDataGroup:2026aaa}. 

In order to gather the data for the fits, we proceed as follows. For those observables for which there is a single measurement, we take the values directly from the experimental papers. For those observables for which there are several measurements, possibly by different experiments, we use the average values given by the Heavy Flavor Averaging Group \cite{HFLAV:2024ctg} or by the Particle Data Group \cite{ParticleDataGroup:2026aaa}.
All the fits are done using the {\it Minuit} package \cite{James:1975dr} to find the minimum $\chi^2$.

\subsection{\boldmath $B \to \rho K^*$ decays, isospin}

As we saw in Sec.~\ref{sec:isospin}, once the EWP-tree relations are taken into account, the four $\btopik$ amplitudes can be expressed in terms of four effective diagrams: $\widetilde{T}$, $\widetilde{C}$ and $\widetilde{P_{uc}}$ (proportional to $\lambda_{u}^{(s)}$), and $\widetilde{P_{tc}}$ (proportional to $\lambda_{t}^{(s)}$). This result also applies to each transversity component of the four $B \to \rho K^*$ decays, making a total of twelve diagrams. This corresponds to 23 unknown parameters, 12 magnitudes of diagrams and 11 relative strong phases. From Appendix \ref{app:data} and the careful counting in Appendix~\ref{app:relations}, we find that there are a total of 26 measured independent observables in the $B \to \rho K^*$ decay modes, so a fit can be done.

As only the differences of strong phases are measurable, in the fit we set the strong phase of the $\widetilde{T}_\perp$ diagram to 0; the strong phases of other diagrams are allowed to vary between 0 and $360^\circ$. The results are shown in Appendix \ref{app:fit_results}: we find an acceptable fit with $\chi^{2}_{\rm min}/\rm d.o.f.$ = \chirhoK/\dofrhoK~ for a $p$-value of \pvalrhoK.

Even so, there is a problem here. As mentioned previously, one can estimate the relative sizes of certain diagrams. In particular, as noted in Sec.~\ref{sec:decomposition}, it is expected that $|P_{uc}/P_{tc}| < 0.5$ [Eq.~(\ref{Puc/Ptc})]. But in Appendix \ref{app:fit_results}, we see that 
\beq
|\widetilde{P_{uc}^0}/\widetilde{P_{tc}^0}| = 229 ~~,~~~~
|\widetilde{P_{uc}^\|}/\widetilde{P_{tc}^\|}| = 164 ~~,~~~~
|\widetilde{P_{uc}^\perp}/\widetilde{P_{tc}^\perp}| = 240 ~.
\eeq
One must be careful, as these are effective diagrams: $\widetilde{P_{uc}} = P_{uc} + A$ and
$\widetilde{P_{tc}} \equiv P_{tc} - \frac{1}{3} (P_{EW}^{P_u}  - P_{EW}^E)$ [Eq.~(\ref{BpiKeffdiags})]. We expect $\widetilde{P_{tc}} \simeq P_{tc}$, but $A$ could be comparable to $P_{uc}$ in $\widetilde{P_{uc}}$. Even so, we expect $|A|$ to be smaller than $|P_{tc}|$, so that the above results exceed expectations [Eq.~(\ref{Puc/Ptc})] by a factor of at least $\sim 300$.

With this in mind, we redo the fit putting a constraint on  $|\widetilde{P_{uc}^\lambda}/\widetilde{P_{tc}^\lambda}|$. Because the ratio $|P_{uc}/P_{tc}|$ as calculated in Ref.~\cite{Buras:1994pb} is a rough estimate, and since there may be other, difficult-to-quantify hadronic effects associated with this ratio, we impose the conservative constraint $|\widetilde{P_{uc}^{\lambda}}/\widetilde{P_{tc}^{\lambda}}| \leq 1$.

The results of the fit with this constraint are shown in Appendix \ref{app:fit_results}. We find that the fit quality is considerably worse: now $\chi^{2}_{\rm min}/\rm d.o.f.$ = \chirhoKConstraint/\dofrhoK, for a $p$-value of \pvalrhoKConstraint. This corresponds to a \sigmarhoKConstraint$\sigma$ discrepancy with the SM, and hints at a potential disagreement between the theoretical expectations and the measurements. The observables that most contribute to the $\chi^2$ in this fit can be found in Appendix \ref{app:contrib_chi2}.
We note that, in the $\btopik$ fits of Refs.~\cite{Berthiaume:2023kmp, Bhattacharya:2025qye, Bhattacharya:2025wcq}, the best-fit value of $|\widetilde{P_{uc}}/\widetilde{P_{tc}}|$ was also larger than expected. There, this ratio was in the range 10-100.

\subsection{\boldmath $B \to VV$ decays, $V \in \{ \rho, K^* \}$, SU(3)$_F$}

We now turn to the combined fit of all the $B \to VV$ modes in which $V \in \{\rho, K^* \}$. Here the final state transforms as $(\boldsymbol{8\otimes8})_S$ under SU(3)$_F$. As we saw in Sec.~\ref{sec:SU3}, once the EWP-tree relations are taken into account, the amplitudes for the $(\boldsymbol{8\otimes8})_S$ $B \to PP$ decays can be expressed in terms of seven effective diagrams: $\widetilde{T}$, $\widetilde{C}$, $\widetilde{P_{uc}}$, $\widetilde{A}$ and $\widetilde{PA_{uc}}$ are proportional to $\lambda_{u}^{(q)}$, while $\widetilde{P_{tc}}$ and $\widetilde{PA_{tc}}$ are proportional to $\lambda_{t}^{(q)}$. This applies to each transversity component of the $B \to VV$ decays, so there are a total of 21 diagrams, corresponding to 41 parameters (21 magnitudes of diagrams and 20 relative strong phases).

As can be seen from the $(\boldsymbol{8\otimes8})_S$ category of Table \ref{tab:BVVdecays}, there are eight $\Delta S = 0$ and eight $\Delta S = 1$ $B \to VV$ decays in which $V \in \{\rho, K^* \}$. However, to date, only six $\Delta S=0$ and five $\Delta S=1$ decays have been measured. The available experimental data in the $\Delta S =0$ and $\Delta S =1$ channels can be found in Appendix \ref{app:data}. Counting only the independent observables, we see that there are 17 (29) observables measured in the $\Delta S=0$ ($\Delta S=1$) modes. This means that, contrary to the fits to the $B \to PP$ data performed in Refs.~\cite{Berthiaume:2023kmp,Bhattacharya:2025wcq}, the individual $\Delta S=0$ and $\Delta S=1$ fits cannot be performed as there are more unknown parameters than there are observables.

On the other hand, the combined $\Delta S=0$ and $\Delta S=1$ fit can be performed with the assumption of SU(3)$_F$ symmetry (i.e., $\Delta S=0$ and $\Delta S=1$ decays are described by the same diagrams). The results are shown in Appendix \ref{app:fit_results}. We find a terrible fit: $\chi^{2}_{\rm min}/{\rm d.o.f.} = \chiVV/\dofVV$. This corresponds to a $p$-value of \pvalVV, or a \sigmaVV$\sigma$ discrepancy with the SU(3)$_F$ limit of the SM. (We note that a sizeable disagreement with the SM was also found in the combined fit of charmless $B \to PP$ decays \cite{Berthiaume:2023kmp,Bhattacharya:2025wcq}.) 
The largest contributions to the $\chi^2$ can be found in Appendix \ref{app:contrib_chi2}.

In fact, the discrepancy with the SM is almost certainly larger than \sigmaVV$\sigma$. Recall that, in Sec.~\ref{sec:decomposition}, we noted that $|P_{uc}/P_{tc}| < 0.5$ is expected [Eq.~(\ref{Puc/Ptc})], but in $B \to \rho K^*$ decays the ratios $|\widetilde{P_{uc}^\lambda}/\widetilde{P_{tc}^\lambda}|$ ($\lambda = 0, \|, \perp$) were larger than this by a factor of at least $\sim 300$. When the constraint $|\widetilde{P_{uc}^{\lambda}}/\widetilde{P_{tc}^{\lambda}}| \leq 1$ was imposed, the $B \to \rho K^*$ fit was found to be in serious disagreement with the expectations of the SM. In the case of the combined $B \to VV$ fit, we see that 
\beq
|\widetilde{P_{uc}^0}/\widetilde{P_{tc}^0}| = 2.2 ~~,~~~~
|\widetilde{P_{uc}^\|}/\widetilde{P_{tc}^\|}| = 4.5 ~~,~~~~
|\widetilde{P_{uc}^\perp}/\widetilde{P_{tc}^\perp}| = 5.2 ~.
\eeq
While these ratios are less extreme than those found in the $B \to \rho K^*$ case, if the fit were redone with the constraint $|\widetilde{P_{uc}^{\lambda}}/\widetilde{P_{tc}^{\lambda}}| \leq 1$ imposed, then one would clearly obtain a larger discrepancy. It is true that this result assumes perfect SU(3)$_F$ symmetry, but even so, it seems clear that there is a serious problem here.

Ideally, one would add SU(3)$_F$-breaking effects in a complete and rigorous way in order to see how large they would have to be to explain the data. However, such a study goes beyond the scope of this analysis. Without this, the best way to estimate the required size of the symmetry breaking is to separate the data into $\Delta S = 0$ and $\Delta S = 1$ decays, perform independent fits, and then compare the central values of the parameters. Unfortunately, since the number of parameters largely surpasses the number of observables, it is not possible to separate the two sectors. An alternative to this decomposition is to add three independent scaling factors (one per polarization) that multiply the parameters for the $\Delta S = 1$ decays. Including these breaking parameters and redoing the $B\to VV$ fit with $V \in \{\rho, K^*\}$, we find that a breaking of 1400\% is required to get an acceptable fit, which we take to be a $p$-value of 5\%. This is largely driven by the breaking in the parallel component, but the breaking in the longitudinal polarization also surpasses expectations, with a required breaking of 500\%. Both of these are far larger than the $\sim 30$\% SU(3)$_F$ breaking expected in the SM.

\subsection{\boldmath $B \to VV$ decays, $V \in \{ \rho, K^*, \phi, \omega \}$, SU(3)$_F$}

The physical states $\phi$ and $\omega$ are admixtures of $\omega_8$ and $\omega_1$, and can be written as 
\begin{align}
\phi &= \omega_8 \, \cos \theta_{w} + \omega_1 \, \sin \theta_{w} ~, \nonumber \\
\omega &=  -\omega_8 \, \sin\theta_{w} + \omega_1 \, \cos \theta_{w} ~, 
\end{align}
where
\begin{align}
\omega_8 \equiv \frac{2 s \bar{s}-u\bar{u}-d\bar{d}}{\sqrt{6}} ~,~~~~
\omega_1  \equiv \frac{u\bar{u}+d\bar{d}+s \bar{s}}{\sqrt{3}} ~.
\end{align}
The ideal mixing angle $\theta_w$ = $\rm arctan(1/\sqrt{2})$ $\approx$ $35.3^\circ$, so that $\phi$ becomes pure $s \bar{s}$ and $\omega$ becomes pure $u \bar{u}+d \bar{d}$.

There are nine $\Delta S=0$ and nine $\Delta S=1$ $B \to VV$ decays in which at least one $V \in \{\phi, \omega \}$. Eight $\Delta S=0$ and six $\Delta S=1$ decays of this type have been measured. The available experimental data for the $\Delta S =0$ and $\Delta S =1$ channels are given in Appendix \ref{app:data}.

As shown above, the physical $\phi$ and $\omega$ mesons are linear combinations of the $\omega_8$ and $\omega_1$ states, which are part of the octet and singlet of SU(3)$_F$. The final states containing a $\phi$ or $\omega$ along with a $\rho$ or $K^*$ meson correspond to the $(\boldsymbol{8\otimes8})_S$ and $\boldsymbol{8\otimes1}$ decays of Table \ref{tab:BVVdecays}, whereas the final $VV$ states in which both $V$s $\in \{\phi, \omega \}$ correspond to all three types of decays of Table \ref{tab:BVVdecays}. 

For the $(\boldsymbol{8\otimes 8})_S$ decays, there are seven independent diagrams for each transversity component. For the $\boldsymbol{8\otimes 1}$ and $\boldsymbol{1\otimes 1}$ decays, there are four and two independent diagrams for each transversity component, respectively. This corresponds to 77 unknown parameters:
39 magnitudes of diagrams and 38 relative strong phases.

As was the case with $B \to VV$ decays in which $V \in \{\rho, K^* \}$, the individual $\Delta S=0$ and $\Delta S=1$ fits cannot be performed as there are more unknown parameters than there are observables. However, the combined $\Delta S=0$ and $\Delta S=1$ fit can be done, and the results are shown in Appendix \ref{app:fit_results}. With the addition of decay modes in which at least one $V \in \{ \phi, \omega \}$, the disagreement with the SM$_{\rm{SU(3)}_F}$ has increased significantly: now $\chi^{2}_{\rm min}/\rm d.o.f. = \chiAll/\dofAll$, for a $p$-value \pvalAll, or a discrepancy of \sigmaAll$\sigma$. The problem we saw in the previous fit is even more pronounced here.

We note that, for the $B \to VV$ fit with $V \in \{ \rho, K^* \}$, we found $\chi^{2}_{\rm min} = $ \chiVV, whereas when $V \in \{ \rho, K^*, \phi, \omega \}$, we have $\chi^{2}_{\rm min} = $ \chiAll. Can we pinpoint the cause of this dramatic increase in $\chi^{2}_{\rm min}$? It turns out to be a variety of factors. Most observables measured in decays with at least one $\phi$ or $\omega$ in the final state do not contribute significantly to $\chi^{2}_{\rm min}$. Those that most contribute can be found in Appendix \ref{app:contrib_chi2}. We see that two observables that have a particularly large $\chi^2$ contribution are ${\cal B}(B^0\to\phi K^{*0})$ and ${\cal B}(B^0\to\omega K^{*0})$, each of which gives a contribution of over 20. The main effect of including in the fit $B \to VV$ decays in which at least one $V \in \{ \phi, \omega \}$ seems to be to increase the tension with the other decays. To be specific, in these new decays, $f_L \ll f_T$ is not observed, in contrast to some decays in which $V \in \{ \rho, K^* \}$. As a result, polarization measurements in $\Delta S =1$ $B \to VV$ decays with no $\phi$ or $\omega$ have larger contributions to $\chi^{2}_{\rm min}$ than before. For example, $\widetilde{f_0}$ in $B^0\to \rho^0 K^{*0}$ now has a contribution of 10, whereas before it contributed 1.9. In fact, the total contribution to $\chi^{2}_{\rm min}$ from polarization measurements in $\Delta S =1$ decays is around 20, compared to around 5 in the fit involving decays with no $\phi$ or $\omega$.

\section{Summary}
\label{sec:conclusion}

Recently, $B\to PP$ decays ($B \in \{B^0, B^+, \bs\}$, $P \in \{ \pi, K, \eta, \eta' \}$) were studied under the assumption of flavor SU(3) symmetry \cite{Berthiaume:2023kmp, Bhattacharya:2025wcq}. Fits to the latest data were performed, and it was found that the individual fits to $\Delta S=0$ or $\Delta S=1$ decays are good. However, the combined fit is very poor: there is a $4.1\sigma$ disagreement with the SU(3)$_F$ limit of the Standard Model. A comparison of the individual $\Delta S=0$ and $\Delta S=1$ results suggests that SU(3)$_F$ is broken at the level of 1000\%, far larger than the $\sim 30\%$ SU(3)$_F$ breaking expected in the SM. 

In this paper, we extend this analysis to $B \to VV$ decays, in which $V \in \{\rho, K^*, \phi, \omega\}$. The methodology is the same as for $B \to PP$ decays, except that the $VV$ final state comes in three helicities. That is, there are three times as many unknown parameters in $B \to VV$ decays as there are in $B \to PP$ decays. But there are also many more $B \to VV$ observables, so that a fit can be done.

Three types of fits have been performed. First, $B \to \rho K^*$ decays are examined, assuming only isospin symmetry. We find that $\chi^{2}_{\rm min}/\rm d.o.f.$ = \chirhoK/\dofrhoK, for a $p$-value of \pvalrhoK. This is an acceptable fit. Second, we study all $B \to VV$ decays in which $V \in \{ \rho, K^* \}$ assuming SU(3)$_F$. There are not enough observables to perform the individual $\Delta S=0$ or $\Delta S=1$ fits, but we can do the combined fit. We find 
$\chi^{2}_{\rm min}/{\rm d.o.f.} = \chiVV/\dofVV$, for  a $p$-value of \pvalVV. This is a terrible fit: it corresponds to a \sigmaVV$\sigma$ discrepancy with the SU(3)$_F$ limit of the SM. Finally, we consider $B \to VV$ decays with $V \in \{ \rho, K^*, \phi, \omega \}$. The fit gives $\chi^{2}_{\rm min}/\rm d.o.f. = \chiAll/\dofAll$, for a $p$-value \pvalAll, or a discrepancy of \sigmaAll$\sigma$ (!).

In these fits the parameter space is very large: there are 23 unknown parameters in the $B \to \rho K^*$ fit, 41 in $B \to VV$ decays with $V \in \{ \rho, K^* \}$, and 77 in the fit to the data for $B \to VV$, $V \in \{ \rho, K^*, \phi, \omega \}$. With so many parameters, how can we be sure that we have found the true minimum $\chi^2$? Given the size of the parameter space, it was necessary to do many crosschecks. We therefore repeated the fits using codes written by three different people, always taking tens of thousands of different starting points. The three fits all found the same results.

The only theoretical input in our analysis is the neglect of the Wilson coefficients $c_7$ and $c_8$. This approximation is expected to be good to about 10\%. Apart from this, our results are rigorous, group theoretically, and hold exactly in the SU(3)$_F$ limit. The only remaining theoretical task is to examine SU(3)$_F$ breaking. It seems unlikely that the introduction of $\sim 30$\% SU(3)$_F$-breaking effects can account for a \sigmaAll$\sigma$ discrepancy with the SM, but this must be checked.

\acknowledgments{We thank D. van Dyk and S. Schacht for helpful comments. D.L. thanks R. Fleischer for helpful communications. This work was financially supported by the National Science Foundation, Grant No.\ PHY-2310627 (B.B.), by NSERC of Canada (M.B., D.L., I.R.), and by FRQNT, Scholarship No.\ 363240 (M.B.).}

\appendix

\section{\boldmath $B \to VV$ amplitudes -- decomposition in terms of RMEs}\label{app:RMEs}

In $B \to VV$ decays, the final-state vector $V$ can be a member of the SU(3)$_F$ octet ${\bf 8} = \{\rho^\pm, \rho^0, K^{*\pm}, K^{*0}, {\bar K}^{*0}, \omega_8 \}$ or the singlet ${\bf 1} = \omega_1$. (The physical $\omega$ and $\phi$ mesons are linear combinations of $\omega_8$ and $\omega_1$, see Sec.~\ref{sec:SU3}.) The $VV$ final state therefore comes in three categories: $(\boldsymbol{8\otimes 8})_S$, $\boldsymbol{8\otimes1}$, or $\boldsymbol{1\otimes1}$. Each category has its own set of RMEs; these are defined in Eqs.~(\ref{eq:8x8RMEs}), (\ref{eq:8x1RMEs}) and (\ref{eq:1x1RMEs}), respectively. All $B \to VV$ decay amplitudes can be decomposed in terms of RMEs. For the three categories of decays, this decomposition is shown below in Tables \ref{tab:8x8RME}, \ref{tab:8x1RME} and \ref{tab:1x1RME}, respectively.

\newpage
\begin{table}[H]
\begin{center}
\makebox[\textwidth]{
\begin{tabular}{|l|l|c|c|c|c|c||c|c|c|c|c|}
         \hline
         \multicolumn{2}{|c|}{\multirow{3}{*}{Decays}} & \multicolumn{5}{c||}{$\lambda_u^{(q)}$} & \multicolumn{5}{c|}{$\lambda_t^{(q)}$}\\
         \cline{3-12}
         \multicolumn{2}{|l|}{}& \multicolumn{5}{c||}{$c_1,c_2$} & \multicolumn{2}{c|}{$c_3$, $c_4$, $c_5$, $c_6$, $c_9$, $c_{10}$} & \multicolumn{3}{c|}{$c_9$, $c_{10}$}\\
         \cline{3-12}
         \multicolumn{2}{|l|}{}& $A_1^u$ & $A_8^u$ & $R_8^u$ & $P_8^u$ & $P_{27}^u$ & $A_1^t$ & $A_8^t$ & $R_8^t$ & $P_8^t$ & $P_{27}^t$\\
         \hline\hline
    
      \multirow{12}{*}{$\Delta S=0$}   & $B^+ \to K^{*+} \bar{K}^{*0}$ & 0 & $-\frac{\sqrt{3}}{\sqrt{5}}$ & $-\frac{1}{\sqrt{5}}$ & $\frac{3\sqrt{3}}{5}$ & $\frac{2\sqrt{3}}{5}$ & 0 & $-\frac{\sqrt{3}}{\sqrt{5}}$ & $\frac{3}{\sqrt{5}}$ & $\frac{9\sqrt{3}}{5}$ & $\frac{6\sqrt{3}}{5}$\\
         &$B^+ \to \rho^+ \rho^0$ & 0 & 0 & 0 & 0 & $\sqrt{6}$ & 0 & 0 & 0 & 0 & $3\sqrt{6}$\\ 
         &$B^+ \to \omega_8 \rho^+$ & 0 & $\frac{\sqrt{2}}{\sqrt{5}}$ & $\frac{\sqrt{2}}{\sqrt{15}}$ & $-\frac{3\sqrt{2}}{5}$ & $\frac{3\sqrt{2}}{5}$&0 & $\frac{\sqrt{2}}{\sqrt{5}}$ & $-\frac{\sqrt{6}}{\sqrt{5}}$ & $-\frac{9\sqrt{2}}{5}$ & $\frac{9\sqrt{2}}{5}$\\
    
         & $B^0 \to K^{*0} \bar{K}^{*0}$& $-\frac{1}{2\sqrt{3}}$ & $-\frac{1}{\sqrt{15}}$ & $-\frac{1}{\sqrt{5}}$ & $-\frac{3\sqrt{3}}{5}$ & $\frac{\sqrt{3}}{10}$ & $-\frac{1}{2\sqrt{3}}$ & $-\frac{1}{\sqrt{15}}$ & $\frac{3}{\sqrt{5}}$ & $-\frac{9\sqrt{3}}{5}$ & $\frac{3\sqrt{3}}{10}$\\
         & $B^0 \to \rho^+ \rho^-$ & $\frac{1}{2\sqrt{3}}$ & $\frac{1}{\sqrt{15}}$ & $-\frac{1}{\sqrt{5}}$ & $-\frac{\sqrt{3}}{5}$ & $\frac{7\sqrt{3}}{10}$ & $\frac{1}{2\sqrt{3}}$ & $\frac{1}{\sqrt{15}}$ & $\frac{3}{\sqrt{5}}$ & $-\frac{3\sqrt{3}}{5}$ & $\frac{21\sqrt{3}}{10}$\\
         &$B^0 \to K^{*+} K^{*-}$ & $\frac{1}{2\sqrt{3}}$ & $-\frac{2}{\sqrt{15}}$ & 0 & $-\frac{2\sqrt{3}}{5}$ & $-\frac{\sqrt{3}}{10}$ & $\frac{1}{2\sqrt{3}}$ & $-\frac{2}{\sqrt{15}}$ & 0 & $-\frac{6\sqrt{3}}{5}$ & $-\frac{3\sqrt{3}}{10}$\\
         &$B^0 \to \rho^0 \rho^0$ & $-\frac{1}{2\sqrt{6}}$ & $-\frac{1}{\sqrt{30}}$ & $\frac{1}{\sqrt{10}}$ & $\frac{\sqrt{3}}{5\sqrt{2}}$ & $\frac{13\sqrt{3}}{10\sqrt{2}}$ & $-\frac{1}{2\sqrt{6}}$ & $-\frac{1}{\sqrt{30}}$ & $-\frac{3}{\sqrt{10}}$ & $\frac{3\sqrt{3}}{5\sqrt{2}}$ & $\frac{39\sqrt{3}}{10\sqrt{2}}$\\
         &$B^0 \to \rho^0 \omega_8$ & 0 & $\frac{1}{\sqrt{5}}$ & $\frac{1}{\sqrt{15}}$ & 1 & 0 & 0 & $\frac{1}{\sqrt{5}}$ & $-\frac{\sqrt{3}}{\sqrt{5}}$ & $3$ & 0\\
          & $B^0 \to  \omega_8 \omega_8$ & $-\frac{1}{2\sqrt{6}}$ & $\frac{1}{\sqrt{30}}$ & $-\frac{1}{\sqrt{10}}$ & $-\frac{\sqrt{3}}{5\sqrt{2}}$ & $-\frac{3\sqrt{3}}{10\sqrt{2}}$ & $-\frac{1}{2\sqrt{6}}$ & $\frac{1}{\sqrt{30}}$ & $\frac{3}{\sqrt{10}}$ & $-\frac{3\sqrt{3}}{5\sqrt{2}}$ & $-\frac{9\sqrt{3}}{10\sqrt{2}}$\\
        
         &$\bs \to \rho^+ K^{*-}$ & 0 & $\frac{\sqrt{3}}{\sqrt{5}}$ & $-\frac{1}{\sqrt{5}}$ & $\frac{\sqrt{3}}{5}$ & $\frac{4\sqrt{3}}{5}$ & 0 & $\frac{\sqrt{3}}{\sqrt{5}}$ & $\frac{3}{\sqrt{5}}$ & $\frac{3\sqrt{3}}{5}$ & $\frac{12\sqrt{3}}{5}$\\
         &$\bs \to \rho^0 \bar{K}^{*0}$ & 0 & $-\frac{\sqrt{3}}{\sqrt{10}}$ & $\frac{1}{\sqrt{10}}$ & $-\frac{\sqrt{3}}{5\sqrt{2}}$ & $\frac{3\sqrt{6}}{5}$ & 0 & $-\frac{\sqrt{3}}{\sqrt{10}}$ & $-\frac{3}{\sqrt{10}}$ & $-\frac{3\sqrt{3}}{5\sqrt{2}}$ & $\frac{9\sqrt{6}}{5}$\\
         &$\bs \to \omega_8 \bar{K}^{*0}$ & 0 & $-\frac{1}{\sqrt{10}}$ & $\frac{1}{\sqrt{30}}$ & $-\frac{1}{5\sqrt{2}}$ & $\frac{3\sqrt{2}}{5}$ & 0 & $-\frac{1}{\sqrt{10}}$ & $-\frac{\sqrt{3}}{\sqrt{10}}$ & $-\frac{3}{5\sqrt{2}}$ & $\frac{9\sqrt{2}}{5}$\\
         \hline\hline
         \multirow{12}{*}{$\Delta S = 1$}
         &$B^+\to \rho^+ K^{*0}$ & 0 & $-\frac{\sqrt{3}}{\sqrt{5}}$ & $-\frac{1}{\sqrt{5}}$ & $\frac{3\sqrt{3}}{5}$ & $\frac{2\sqrt{3}}{5}$ & 0 & $-\frac{\sqrt{3}}{\sqrt{5}}$ & $\frac{3}{\sqrt{5}}$ & $\frac{9\sqrt{3}}{5}$ & $\frac{6\sqrt{3}}{5}$\\
         &$B^+ \to \rho^0 K^{*+}$ & 0 & $\frac{\sqrt{3}}{\sqrt{10}}$ &$\frac{1}{\sqrt{10}}$ & $-\frac{3\sqrt{3}}{5\sqrt{2}}$ & $\frac{4\sqrt{6}}{5}$ & 0 & $\frac{\sqrt{3}}{\sqrt{10}}$ & $-\frac{3}{\sqrt{10}}$ & $-\frac{9\sqrt{3}}{5\sqrt{2}}$ & $\frac{12\sqrt{6}}{5}$\\
          & $B^+ \to \omega_8 K^{*+}$ & 0 & $-\frac{1}{\sqrt{10}}$ & $-\frac{1}{\sqrt{30}}$ & $\frac{3}{5\sqrt{2}}$ & $\frac{6\sqrt{2}}{5}$ & 0 & $-\frac{1}{\sqrt{10}}$ & $\frac{\sqrt{3}}{\sqrt{10}}$ & $\frac{9}{5\sqrt{2}}$ & $\frac{18\sqrt{2}}{5}$ \\
        
         &$B^0 \to \rho^-K^{*+}$ & 0 & $\frac{\sqrt{3}}{\sqrt{5}}$ & $-\frac{1}{\sqrt{5}}$ & $\frac{\sqrt{3}}{5}$ & $\frac{4\sqrt{3}}{5}$ & 0 & $\frac{\sqrt{3}}{\sqrt{5}}$ & $\frac{3}{\sqrt{5}}$ & $\frac{3\sqrt{3}}{5}$ & $\frac{12\sqrt{3}}{5}$\\
         &$B^0 \to \rho^0 K^{*0}$ & 0 & $-\frac{\sqrt{3}}{\sqrt{10}}$ & $\frac{1}{\sqrt{10}}$ & $-\frac{\sqrt{3}}{5\sqrt{2}}$ & $\frac{3\sqrt{6}}{5}$ & 0 & $-\frac{\sqrt{3}}{\sqrt{10}}$ & $-\frac{3}{\sqrt{10}}$ & $-\frac{3\sqrt{3}}{5\sqrt{2}}$ & $\frac{9\sqrt{6}}{5}$\\
         &$B^0 \to \omega_8 K^{*0}$ & 0 & $-\frac{1}{\sqrt{10}}$ & $\frac{1}{\sqrt{30}}$ & $-\frac{1}{5\sqrt{2}}$ & $\frac{3\sqrt{2}}{5}$ & 0 & $-\frac{1}{\sqrt{10}}$ & $-\frac{\sqrt{3}}{\sqrt{10}}$ & $-\frac{3}{5\sqrt{2}}$ & $\frac{9\sqrt{2}}{5}$\\
    
         &$\bs \to K^{*0} \bar{K}^{*0}$ & $-\frac{1}{2\sqrt{3}}$ & $-\frac{1}{\sqrt{15}}$ & $-\frac{1}{\sqrt{5}}$ & $-\frac{3\sqrt{3}}{5}$ & $\frac{\sqrt{3}}{10}$ & $-\frac{1}{2\sqrt{3}}$ & $-\frac{1}{\sqrt{15}}$ & $\frac{3}{\sqrt{5}}$ & $-\frac{9\sqrt{3}}{5}$ & $\frac{3\sqrt{3}}{10}$\\
         &$\bs \to \rho^+ \rho^-$ & $\frac{1}{2\sqrt{3}}$ & $-\frac{2}{\sqrt{15}}$ & 0 & $-\frac{2\sqrt{3}}{5}$ & $-\frac{\sqrt{3}}{10}$ & $\frac{1}{2\sqrt{3}}$ & $-\frac{2}{\sqrt{15}}$ & 0 & $-\frac{6\sqrt{3}}{5}$ & $-\frac{3\sqrt{3}}{10}$\\
         &$\bs \to K^{*+} K^{*-}$ & $\frac{1}{2\sqrt{3}}$ & $\frac{1}{\sqrt{15}}$ & $-\frac{1}{\sqrt{5}}$ & $-\frac{\sqrt{3}}{5}$ & $\frac{7\sqrt{3}}{10}$ & $\frac{1}{2\sqrt{3}}$ & $\frac{1}{\sqrt{15}}$ & $\frac{3}{\sqrt{5}}$ & $-\frac{3\sqrt{3}}{5}$ & $\frac{21\sqrt{3}}{10}$\\
         &$\bs \to \rho^0 \rho^0$ & $-\frac{1}{2\sqrt{6}}$ & $\frac{\sqrt{2}}{\sqrt{15}}$ & 0 & $\frac{\sqrt{6}}{5}$ & $\frac{\sqrt{3}}{10\sqrt{2}}$ & $-\frac{1}{2\sqrt{6}}$ & $\frac{\sqrt{2}}{\sqrt{15}}$ & 0 & $\frac{3\sqrt{6}}{5}$ & $\frac{3\sqrt{3}}{10\sqrt{2}}$\\
         &$\bs \to \rho^0 \omega_8$ & 0 & 0 & $\frac{2}{\sqrt{15}}$ & $\frac{4}{5}$ & $\frac{6}{5}$ & 0 & 0 & $-\frac{2\sqrt{3}}{\sqrt{5}}$ & $\frac{12}{5}$ & $\frac{18}{5}$\\
         &$\bs \to \omega_8 \omega_8$ & $-\frac{1}{2\sqrt{6}}$ & $-\frac{\sqrt{2}}{\sqrt{15}}$ & 0 & $-\frac{\sqrt{6}}{5}$ & $\frac{9\sqrt{3}}{10\sqrt{2}}$ & $-\frac{1}{2\sqrt{6}}$ & $-\frac{\sqrt{2}}{\sqrt{15}}$ & 0 & $-\frac{3\sqrt{6}}{5}$ & $\frac{27\sqrt{3}}{10\sqrt{2}}$\\
         \hline
    \end{tabular}}
\end{center}
\caption{SU(3)$_F$ RME contributions to the $\Delta S = 0$ and $\Delta S=1$ $B \to VV$ amplitudes for the $(\boldsymbol{8\otimes 8})_S$ final states. These apply to each polarization of the $B \to VV$ decays. }
\label{tab:8x8RME}
\end{table}

\begin{table}[H]
\begin{center}
    \begin{tabular}{|l|l|c|c|c||c|c|c|}
         \hline
         \multicolumn{2}{|c|}{\multirow{3}{*}{Decays}} & \multicolumn{3}{c||}{$\lambda_u^{(q)}$} &\multicolumn{3}{c|}{$\lambda_t^{(q)}$}  \\
         \cline{3-8}
         \multicolumn{2}{|c|}{}&\multicolumn{3}{c||}{$c_1,c_2$}&$c_3$, $c_4$, $c_5$, $c_6$, $c_9$, $c_{10}$&\multicolumn{2}{c|}{$c_9$, $c_{10}$}\\
         \cline{3-8}
         \multicolumn{2}{|c|}{}& $C_8^u$ & $L_8^u$ & $M_8^u$ & $C_8^t$ & $L_8^t$ & $M_8^t$\\
         \hline\hline
         \multirow{4}{*}{$\Delta S = 0$}
         &$B^+ \to \rho^+ \omega_1$  & 1 & $\frac{1}{\sqrt{3}}$ & $-\frac{3}{\sqrt{5}}$ & $1$ & $-\sqrt{3}$ & $-\frac{9}{\sqrt{5}}$\\
         \cline{2-8}
         &$B^0 \to \rho^0 \omega_1$  & $\frac{1}{\sqrt{2}}$ & $\frac{1}{\sqrt{6}}$ & $\frac{\sqrt{5}}{\sqrt{2}}$ & $\frac{1}{\sqrt{2}}$ & $-\frac{\sqrt{3}}{\sqrt{2}}$ & $\frac{3\sqrt{5}}{\sqrt{2}}$\\
         &$B^0 \to \omega_1 \omega_8$  & $-\frac{1}{\sqrt{6}}$ & $\frac{1}{\sqrt{2}}$ & $\frac{\sqrt{3}}{\sqrt{10}}$ & $-\frac{1}{\sqrt{6}}$ & $-\frac{3}{\sqrt{2}}$ & $\frac{3\sqrt{3}}{\sqrt{10}}$\\
         \cline{2-8}
         &$\bs \to \omega_1 \bar{K}^{*0}$  & 1 & $-\frac{1}{\sqrt{3}}$ & $\frac{1}{\sqrt{5}}$ & $1$ & $\sqrt{3}$ & $\frac{3}{\sqrt{5}}$\\
         \hline\hline
         \multirow{4}{*}{$\Delta S=1$}
         &$B^+ \to \omega_1 K^{*+}$ & 1 & $\frac{1}{\sqrt{3}}$ & $-\frac{3}{\sqrt{5}}$ & $1$ & $-\sqrt{3}$ & $-\frac{9}{\sqrt{5}}$ \\
         \cline{2-8}
         &$B^0 \to \omega_1 K^{*0}$ & 1 & $-\frac{1}{\sqrt{3}}$ & $\frac{1}{\sqrt{5}}$ & $1$ & $\sqrt{3}$ & $\frac{3}{\sqrt{5}}$\\
         \cline{2-8}
         &$\bs \to \rho^0 \omega_1$ & 0 & $\frac{\sqrt{2}}{\sqrt{3}}$ & $\frac{2\sqrt{2}}{\sqrt{5}}$ & 0 & $-\sqrt{6}$ & $\frac{6\sqrt{2}}{\sqrt{5}}$\\
         &$\bs \to \omega_1 \omega_8$ & $\frac{\sqrt{2}}{\sqrt{3}}$ & 0 & $\frac{\sqrt{6}}{\sqrt{5}}$ & $\frac{\sqrt{2}}{\sqrt{3}}$ & 0 & $\frac{3\sqrt{6}}{\sqrt{5}}$\\
         \hline
    \end{tabular}
\end{center}    
\caption{RME contributions to the $\Delta S = 0$ and $\Delta S=1$ $B \to VV$ amplitudes for the $\boldsymbol{8\otimes 1}$ final states. These apply to each polarization of the $B \to VV$ decays. 
\label{tab:8x1RME}}
\end{table}

\begin{table}[H]
\begin{center}    
\begin{tabular}{|l|l|c||c|} \hline
        \multicolumn{2}{|c|}{\multirow{2}{*}{Decays}} & $\lambda_u^{(q)}$ & $\lambda_t^{(q)}$\\
        \cline{3-4}
        \multicolumn{2}{|c|}{} &$C_1^u$ & $C_1^t$\\
        \hline\hline
        $\Delta S =0$ & $B^0 \to \omega_1 \omega_1$ &$\frac{1}{\sqrt{3}}$ & $\frac{1}{\sqrt{3}}$ \\
        \hline\hline
        $\Delta S=1$ &$\bs \to \omega_1 \omega_1$ & $\frac{1}{\sqrt{3}}$ & $\frac{1}{\sqrt{3}}$\\
        \hline
\end{tabular}
\end{center}
\caption{RME contributions to the $\Delta S = 0$ and $\Delta S=1$ $B \to VV$ amplitudes for the $\boldsymbol{1\otimes 1}$ final state. These apply to each polarization of the $B \to VV$ decays. 
\label{tab:1x1RME}}
\end{table}

\section{\boldmath $B \to VV$ amplitudes -- decomposition in terms of diagrams}
\label{app:diags}

The $B \to VV$ decay amplitudes can also be decomposed in terms of diagrams. For the three categories of decays, $(\boldsymbol{8\otimes 8})_S$, $\boldsymbol{8\otimes1}$, or $\boldsymbol{1\otimes1}$, this decomposition is shown below in Tables \ref{tab:8x8dia}, \ref{tab:8x1dia} and \ref{tab:1x1dia}, respectively. Note that, even though we use the same symbols for the diagrams in the $(\boldsymbol{8\otimes 8})_S$, $\boldsymbol{8\otimes 1}$ and $\boldsymbol{1\otimes 1}$ decay amplitudes ($T$, $C$, etc.), they are different in the three  decay categories.

\begin{table}[H]
\begin{center}
    \centering
    \setlength{\tabcolsep}{0.8pt}
\makebox[\textwidth]{
\begin{tabular}{|l|l|c|c|c|c|c|c||c|c|c|c|c|c|c|c|}
         \hline
         \multicolumn{2}{|c}{\multirow{2}{*}{Decays}} & \multicolumn{6}{|c||}{$\lambda_u^{(q)}$} & \multicolumn{8}{c|}{$\lambda_t^{(q)}$}\\
         \cline{3-16}
         \multicolumn{2}{|c|}{}& $T$ & $C$ & $P_{uc}$ & $A$ & $PA_{uc}$ & $E$ & $P_{tc}$ & $PA_{tc}$ & $P_{EW}$ & $P^C_{EW}$ & $P^A_{EW}$ & $P^E_{EW}$ & $P^{P_{u}}_{EW}$ & $P^{PA_{u}}_{EW}$ \\
         \hline\hline
         \multirow{12}{*}{$\Delta S = 0$}& $B^+ \to K^{*+} \bar{K}^{*0}$ & 0 & 0 & 1 & 1 & 0 & 0 & 1 & 0 & 0 & $-\frac{1}{3}$ & 0 & $\frac{2}{3}$ & $-\frac{1}{3}$ & 0\\
         &$B^+ \to \rho^+ \rho^0$ & $-\frac{1}{\sqrt{2}}$ & $-\frac{1}{\sqrt{2}}$ & 0 & 0 & 0 & 0 & 0 & 0 & $-\frac{1}{\sqrt{2}}$ & $-\frac{1}{\sqrt{2}}$ & 0 & 0 & 0 & 0\\
          &$B^+ \to \omega_8 \rho^+$ & $-\frac{1}{\sqrt{6}}$ & $-\frac{1}{\sqrt{6}}$ & $-\frac{2}{\sqrt{6}}$ & $-\frac{2}{\sqrt{6}}$ & 0 & 0 & $-\frac{2}{\sqrt{6}}$ & 0 & $-\frac{1}{\sqrt{6}}$ & $-\frac{1}{3\sqrt{6}}$ & 0 & $-\frac{4}{3\sqrt{6}}$ & $\frac{2}{3\sqrt{6}}$ & 0\\
         &$B^0 \to K^{*0} \bar{K}^{*0}$ & 0 & 0 & 1 & 0 & 1 & 0 & 1 & 1 & 0 & $-\frac{1}{3}$ & $-\frac{2}{3}$ & $-\frac{1}{3}$ & $-\frac{1}{3}$ & $-\frac{1}{3}$\\
         &$B^0 \to \rho^+ \rho^-$ & $-1$ & 0 & $-1$ & 0 & $-1$ & $-1$ & $-1$ & $-1$ & 0 & $-\frac{2}{3}$ & $-\frac{1}{3}$ & $\frac{1}{3}$ & $\frac{1}{3}$ & $\frac{1}{3}$\\
         &$B^0 \to K^{*+} K^{*-}$ & 0 & 0 & 0 & 0 & $-1$ & $-1$ & 0 & $-1$ & 0 & 0 & $-\frac{1}{3}$ & 0 & 0 & $\frac{1}{3}$\\
         &$B^0 \to \rho^0 \rho^0$ & 0 & $-\frac{1}{\sqrt{2}}$ & $\frac{1}{\sqrt{2}}$ & 0 & $\frac{1}{\sqrt{2}}$ & $\frac{1}{\sqrt{2}}$ & $\frac{1}{\sqrt{2}}$ & $\frac{1}{\sqrt{2}}$ & $-\frac{1}{\sqrt{2}}$ & $-\frac{1}{3\sqrt{2}}$ & $\frac{1}{3\sqrt{2}}$ & $-\frac{1}{3\sqrt{2}}$ & $-\frac{1}{3\sqrt{2}}$ & $-\frac{1}{3\sqrt{2}}$\\
          &$B^0 \to \rho^0 \omega_8$ & 0 & 0 & $-\frac{1}{\sqrt{3}}$ & 0 & 0 & $\frac{1}{\sqrt{3}}$ & $-\frac{1}{\sqrt{3}}$ & 0 & 0 & $\frac{1}{3\sqrt{3}}$ & $\frac{1}{\sqrt{3}}$ & $\frac{1}{3\sqrt{3}}$ & $\frac{1}{3\sqrt{3}}$ & 0 \\
          &$B^0 \to \omega_8 \omega_8$ & 0 & $\frac{1}{3\sqrt{2}}$ & $\frac{1}{3\sqrt{2}}$ & 0 & $\frac{1}{\sqrt{2}}$ & $\frac{1}{3\sqrt{2}}$ & $\frac{1}{3\sqrt{2}}$ & $\frac{1}{\sqrt{2}}$ & $\frac{1}{3\sqrt{2}}$ & $-\frac{1}{9\sqrt{2}}$ & $-\frac{1}{3\sqrt{2}}$ & $-\frac{1}{9\sqrt{2}}$ & $-\frac{1}{9\sqrt{2}}$ & $-\frac{1}{3\sqrt{2}}$\\
         &$\bs \to \rho^+ K^{*-}$ & $-1$ & 0 & $-1$ & 0 & 0 & 0 & $-1$ & 0 & 0 & $-\frac{2}{3}$ & 0 & $\frac{1}{3}$ & $\frac{1}{3}$ & 0\\
         &$\bs \to \rho^0 \bar{K}^{*0}$ & 0  & $-\frac{1}{\sqrt{2}}$ & $\frac{1}{\sqrt{2}}$ & 0 & 0 & 0 & $\frac{1}{\sqrt{2}}$ & 0 & $-\frac{1}{\sqrt{2}}$ & $-\frac{1}{3\sqrt{2}}$ & 0 & $-\frac{1}{3\sqrt{2}}$ & $-\frac{1}{3\sqrt{2}}$ & 0\\
          &$\bs \to \omega_8 \bar{K}^{*0}$ & 0 & $-\frac{1}{\sqrt{6}}$ & $\frac{1}{\sqrt{6}}$ & 0 & 0 & 0 & $\frac{1}{\sqrt{6}}$ & 0 & $-\frac{1}{\sqrt{6}}$ & $-\frac{1}{3\sqrt{6}}$ & 0 & $-\frac{1}{3\sqrt{6}}$ & $-\frac{1}{3\sqrt{6}}$ & 0\\
         \hline\hline
         \multirow{12}{*}{$\Delta S = 1$}
         &$B^+ \to \rho^+ K^{*0}$ & 0 & 0 & 1 & 1 & 0 & 0 & 1 & 0 & 0 & $-\frac{1}{3}$ & 0 & $\frac{2}{3}$ & $-\frac{1}{3}$ & 0\\
         &$B^+ \to \rho^0 K^{*+}$ & $-\frac{1}{\sqrt{2}}$ & $-\frac{1}{\sqrt{2}}$ & $-\frac{1}{\sqrt{2}}$ &  $-\frac{1}{\sqrt{2}}$ & 0 & 0 &$-\frac{1}{\sqrt{2}}$ & 0 & $-\frac{1}{\sqrt{2}}$ & $-\frac{\sqrt{2}}{3}$ & 0 & $-\frac{\sqrt{2}}{3}$ & $\frac{1}{3\sqrt{2}}$ & 0\\
          &$B^+ \to \omega_8 K^{*+}$ & $-\frac{1}{\sqrt{6}}$ & $-\frac{1}{\sqrt{6}}$ & $\frac{1}{\sqrt{6}}$ & $\frac{1}{\sqrt{6}}$ & 0 & 0 & $\frac{1}{\sqrt{6}}$ & 0 & $-\frac{1}{\sqrt{6}}$ & $-\frac{4}{3\sqrt{6}}$ & 0 & $\frac{2}{3\sqrt{6}}$ & $-\frac{1}{3\sqrt{6}}$ & 0\\
         &$B^0 \to \rho^-K^{*+}$ & $-1$ & 0 & $-1$ & 0 & 0 & 0 & $-1$ & 0 & 0 & $-\frac{2}{3}$ & 0 & $\frac{1}{3}$ & $\frac{1}{3}$ & 0\\
         &$B^0 \to \rho^0 K^{*0}$ & 0 & $-\frac{1}{\sqrt{2}}$ & $\frac{1}{\sqrt{2}}$ &  0 & 0 & 0 & $\frac{1}{\sqrt{2}}$ & 0 & $-\frac{1}{\sqrt{2}}$ & $-\frac{1}{3\sqrt{2}}$ & 0 & $-\frac{1}{3\sqrt{2}}$ & $-\frac{1}{3\sqrt{2}}$ & 0\\
          &$B^0 \to \omega_8 K^{*0}$ & 0 & $-\frac{1}{\sqrt{6}}$ & $\frac{1}{\sqrt{6}}$ &  0 & 0 & 0 & $\frac{1}{\sqrt{6}}$ & 0 & $-\frac{1}{\sqrt{6}}$ & $-\frac{1}{3\sqrt{6}}$ & 0 & $-\frac{1}{3\sqrt{6}}$ & $-\frac{1}{3\sqrt{6}}$ & 0\\
         &$\bs \to K^{*0} \bar{K}^{*0}$ & 0 & 0 & 1 & 0 & 1 & 0 & 1 & 1 & 0 & $-\frac{1}{3}$ & $-\frac{2}{3}$ & $-\frac{1}{3}$ & $-\frac{1}{3}$ & $-\frac{1}{3}$\\
         &$\bs \to \rho^+ \rho^-$ & 0 & 0 & 0 & 0 & $-1$ & $-1$ & 0 & $-1$ & 0 & 0 & $-\frac{1}{3}$ & 0 & 0 & $\frac{1}{3}$\\
         &$\bs \to K^{*+} K^{*-}$ & $-1$ & 0 & $-1$ & 0 & $-1$ & $-1$ & $-1$ & $-1$ & 0 & $-\frac{2}{3}$ & $-\frac{1}{3}$ & $\frac{1}{3}$ & $\frac{1}{3}$ & $\frac{1}{3}$\\
         &$\bs \to \rho^0 \rho^0$ & 0 & 0 & 0 & 0 & $\frac{1}{\sqrt{2}}$ & $\frac{1}{\sqrt{2}}$ & 0 & $\frac{1}{\sqrt{2}}$ & 0 & 0 & $\frac{1}{3\sqrt{2}}$ & 0 & 0 & $-\frac{1}{3\sqrt{2}}$\\
          &$\bs \to \rho^0 \omega_8$ & 0 & $-\frac{1}{\sqrt{3}}$ & 0 & 0 & 0 & $\frac{1}{\sqrt{3}}$ & 0 & 0 & $-\frac{1}{\sqrt{3}}$ & 0 & $\frac{1}{\sqrt{3}}$ & 0 & 0 & 0\\
          &$\bs \to \omega_8 \omega_8$ & 0 & $-\frac{\sqrt{2}}{3}$ & $\frac{2\sqrt{2}}{3}$ & 0 & $\frac{1}{\sqrt{2}}$ & $\frac{1}{3\sqrt{2}}$ & $\frac{2\sqrt{2}}{3}$ & $\frac{1}{\sqrt{2}}$ & $-\frac{\sqrt{2}}{3}$ & $-\frac{2\sqrt{2}}{9}$ & $-\frac{1}{3\sqrt{2}}$ & $-\frac{2\sqrt{2}}{9}$ & $-\frac{2\sqrt{2}}{9}$ & $-\frac{1}{3\sqrt{2}}$\\
         \hline
    \end{tabular}}
\end{center}
\caption{Diagrammatic contributions to the $\Delta S = 0$ and $\Delta S=1$ $B \to VV$ amplitudes for the $(\boldsymbol{8\otimes 8})_S$ final states. These apply to each polarization of the $B \to VV$ decays.}
\label{tab:8x8dia}
\end{table}

\begin{table}[H]
\begin{center}
\setlength{\tabcolsep}{0.8pt}
\begin{tabular}{|l|l|c|c|c|c|c|c||c|c|c|c|c|c|c|c|}
         \hline
         \multicolumn{2}{|c|}{\multirow{2}{*}{Decays}} & \multicolumn{6}{c||}{$\lambda_u^{(q)}$} & \multicolumn{8}{c|}{$\lambda_t^{(q)}$}\\
         \cline{3-16}\multicolumn{2}{|c|}{}& $T$ & $C$ & $P_{uc}$ & $A$ & $PA_{uc}$ & $E$ & $P_{tc}$ & $PA_{tc}$ & $P_{EW}$ & $P^C_{EW}$ & $P^A_{EW}$ & $P^E_{EW}$ & $P^{P_{u}}_{EW}$ & $P^{PA_{u}}_{EW}$\\
         \hline\hline
         \multirow{4}{*}{$\Delta S=0$}
         &$B^+ \to \rho^+ \omega_1$ & $\frac{1}{\sqrt{3}}$ & $\frac{1}{\sqrt{3}}$ & $\frac{2}{\sqrt{3}}$ & $\frac{2}{\sqrt{3}}$ & 0 & 0 &$\frac{2}{\sqrt{3}}$ & 0 & 0 & $\frac{1}{3\sqrt{3}}$ & 0 & $\frac{4}{3\sqrt{3}}$ & $-\frac{2}{3\sqrt{3}}$ & 0 \\
         \cline{2-16}
         &$B^0 \to \rho^0 \omega_1$ & 0 & 0 & $\frac{2}{\sqrt{6}}$ & 0 & 0 &$-\frac{2}{\sqrt{6}}$ & $\frac{2}{\sqrt{6}}$ & 0 & $-\frac{1}{\sqrt{6}}$ & $-\frac{2}{3\sqrt{6}}$ & $-\frac{2}{\sqrt{6}}$ & $-\frac{2}{3\sqrt{6}}$ & $-\frac{2}{3\sqrt{6}}$ & 0 \\
         &$B^0 \to \omega_1 \omega_8$ & 0 & $-\frac{\sqrt{2}}{3}$ & $-\frac{\sqrt{2}}{3}$ & 0 & 0 & $-\frac{\sqrt{2}}{3}$ & $-\frac{\sqrt{2}}{3}$ & 0 & $-\frac{1}{3\sqrt{2}}$ & $\frac{\sqrt{2}}{9}$ & $-\frac{\sqrt{2}}{3}$ & $\frac{\sqrt{2}}{9}$ & $\frac{\sqrt{2}}{9}$ & 0 \\
         \cline{2-16}
         &$\bs \to \omega_1 \bar{K}^{*0}$ & 0 & $\frac{1}{\sqrt{3}}$ & $\frac{2}{\sqrt{3}}$ & 0 & 0 & 0 & $\frac{2}{\sqrt{3}}$ & 0 & 0 & $-\frac{2}{3\sqrt{3}}$ & 0 & $-\frac{2}{3\sqrt{3}}$ & $-\frac{2}{3\sqrt{3}}$ & 0 \\
         \hline\hline
         \multirow{4}{*}{$\Delta S=1$}
         & $B^+ \to \omega_1 K^{*+}$ & $\frac{1}{\sqrt{3}}$ & $\frac{1}{\sqrt{3}}$ & $\frac{2}{\sqrt{3}}$ & $\frac{2}{\sqrt{3}}$ & 0 & 0 & $\frac{2}{\sqrt{3}}$ & 0 & 0 & $\frac{1}{3\sqrt{3}}$ & 0 & $\frac{4}{3\sqrt{3}}$ & $-\frac{2}{3\sqrt{3}}$ & 0 \\
         \cline{2-16}
         & $B^0 \to \omega_1 K^{*0}$ & 0 & $\frac{1}{\sqrt{3}}$ & $\frac{2}{\sqrt{3}}$ & 0 & 0 & 0 & $\frac{2}{\sqrt{3}}$ & 0 & 0 & $-\frac{2}{3\sqrt{3}}$ & 0 & $-\frac{2}{3\sqrt{3}}$ & $-\frac{2}{3\sqrt{3}}$ & 0 \\
         \cline{2-16}
         & $\bs \to \rho^0 \omega_1$ & 0 & $-\frac{1}{\sqrt{6}}$ & 0 & 0 & 0 & $-\frac{2}{\sqrt{6}}$ & 0 & 0 & $-\frac{1}{\sqrt{6}}$ & 0 & $-\frac{2}{\sqrt{6}}$ & 0 & 0 & 0\\
         &$\bs \to \omega_8 \omega_1$ & 0 & $\frac{1}{3\sqrt{2}}$ & $\frac{2\sqrt{2}}{3}$ & 0 & 0 & $-\frac{2}{3\sqrt{2}}$ & $\frac{2\sqrt{2}}{3}$ & 0 & $-\frac{1}{3\sqrt{2}}$ & $-\frac{2\sqrt{2}}{9}$ & $-\frac{\sqrt{2}}{3}$ & $-\frac{2\sqrt{2}}{9}$ & $-\frac{2\sqrt{2}}{9}$ & 0 \\
         \hline
\end{tabular}
\end{center}
\caption{Diagrammatic contributions to the $\Delta S = 0$ and $\Delta S=1$ $B \to VV$ amplitudes for the $\boldsymbol{8\otimes 1}$ final states. These apply to each polarization of the $B \to VV$ decays. \label{tab:8x1dia}}
\end{table}

\begin{table}[H]
\begin{center}
    \begin{tabular}{|l|l|c|c|c|c|c|c||c|c|c|c|c|c|c|c|}
        \hline
        \multicolumn{2}{|c|}{\multirow{2}{*}{Decays}} & \multicolumn{6}{c||}{$\lambda_u^{(q)}$} & \multicolumn{8}{c|}{$\lambda_t^{(q)}$}\\
        \cline{3-16}
        \multicolumn{2}{|c|}{}& $T$ & $C$ & $P_{uc}$ & $A$ & $PA_{uc}$ & $E$ & $P_{tc}$ & $PA_{tc}$ & $P_{EW}$ & $P^C_{EW}$ & $P^A_{EW}$ & $P^E_{EW}$ & $P^{P_{u}}_{EW}$ & $P^{PA_{u}}_{EW}$ \\
        \hline\hline
        $\Delta S = 0$ & $B^0 \to \omega_1 \omega_1$ & 0 & $\frac{\sqrt{2}}{3}$ & $\frac{\sqrt{2}}{3}$ & 0 & $2\sqrt{2}$ & $\frac{\sqrt{2}}{3}$ & $\frac{\sqrt{2}}{3}$ & $2\sqrt{2}$ & 0 & $-\frac{\sqrt{2}}{9}$ & 0 & $-\frac{\sqrt{2}}{9}$ & $-\frac{\sqrt{2}}{9}$ & $-\frac{2\sqrt{2}}{3}$\\
        \hline\hline
        $\Delta S = 1$ & $\bs \to \omega_1 \omega_1$ & 0 & $\frac{\sqrt{2}}{3}$ & $\frac{\sqrt{2}}{3}$ & 0 & $2\sqrt{2}$ & $\frac{\sqrt{2}}{3}$ & $\frac{\sqrt{2}}{3}$ & $2\sqrt{2}$ & 0 & $-\frac{\sqrt{2}}{9}$ & 0 & $-\frac{\sqrt{2}}{9}$ & $-\frac{\sqrt{2}}{9}$ & $-\frac{2\sqrt{2}}{3}$\\
        \hline
    \end{tabular}
\end{center}
\caption{Diagrammatic contributions to the $\Delta S = 0$ and $\Delta S=1$ $B \to VV$ amplitudes for the $\boldsymbol{1\otimes 1}$ final states. These apply to each polarization of the $B \to VV$ decays. \label{tab:1x1dia}}
\end{table}

\section{\boldmath Measured $B \to VV$ observables}
\label{app:data}
Tables \ref{tab:dataS0}- \ref{tab:dataS1omega} present the data used in our analysis. The data is separated between $\Delta S=0$ and $\Delta S = 1$ decays, and between the decays that have no $\omega$ or $\phi$ in the final state and those that have at least one of these particles. The different observables are described in Sec.~\ref{sec:observables}.

\begin{table} [H]
\centering
\makebox[\textwidth]{
\begin{tabular}{|c|c|c|c|c|c|c|}
\hline
Decays  & $B^+ \to K^{*+} \bar{K}^{*0}$ &$B^+ \to \rho^+ \rho^0$ & $B^0 \to K^{*0} \bar{K}^{*0}$  & $B^0 \to \rho^+ \rho^-$& $B^0\to K^{*+} K^{*-}$& $B^0 \to \rho^0 \rho^0$    \\
\hline
${\cal B}_{CP}$ ($\times 10^{-6}$)   & 0.91 $\pm$ 0.30 & $24.0 \pm 1.9$ &  $0.68^{+0.15}_{-0.14}$ & $28.0 \pm 1.7$ & $0.52^{+0.83\dagger}_{-0.58}$& $0.96 \pm 0.15$\\
${\cal A}_{CP}$ & & $-0.051\pm0.054$ &   & $0.00\pm0.08^*$ & &  $-0.2\pm0.9^{\dagger\dagger}$  \\
${\cal S}_{CP}$ &  &  & & $-0.18\pm0.11^*$&  & $0.3\pm0.7^{\dagger\dagger}$  \\
$\widetilde{f}_{0}$ & $0.82^{+0.13}_{-0.17}$ & $0.950\pm0.016$ & $0.61\pm0.03$ & $0.970^{+0.015}_{-0.016}$ & & $0.71\pm0.06$   \\
$\widetilde{f}_{||}$    & & & $0.170\pm0.028$  & & &\\
$\widetilde{f}_{\perp}$ & & & $0.240\pm0.028$ & & & \\
\hline 
\end{tabular}}
\caption{Observables for $\Delta S=0$ $B \to VV$ decays in which $V \in \{\rho, K^* \}$. Data labeled $^*$, $^{\dagger}$, and $^{\dagger\dagger}$ are taken from Refs.~\cite{ParticleDataGroup:2026aaa}, \cite{BaBar:2008kow}, and \cite{BaBar:2008xku}, respectively. The remaining data are taken from Ref.~\cite{HFLAV:2024ctg}.}
\label{tab:dataS0}
\end{table}

\begin{table} [H]
\centering
\makebox[\textwidth]{
\begin{tabular}{|c|c|c|c|c|c|}
\hline
Observables  &  $B^+ \to \rho^+ K^{*0}$ & $B^+ \to \rho^0 K^{*+}$ & $B^0 \to \rho^- K^{*+}$ & $B^0 \to \rho^0 K^{*0}$ & $\bs \to K^{*0} \bar{K}^{*0}$   \\
\hline
${\cal B}_{CP}$ ($\times 10^{-6}$) & 9.2 $\pm$ $1.5$ & 4.6 $\pm$ $1.1$ & 10.3 $\pm$ $2.6$ & 3.88 $\pm$ $0.77$ &  9.38 $\pm$ $0.48^{\dagger \dagger\dagger}$  \\
${\cal A}_{CP}$  & $-$0.01 $\pm$ $0.16$ & $0.470\pm0.058$ & 0.21 $\pm$ $0.15$ & $-$0.06 $\pm$ $0.09$ & \\
$\widetilde{f}_{0}$ & 0.48 $\pm$ $0.08$ & $0.66 \pm 0.04^*$ & 0.38  $\pm$ $0.13$ & 0.173 $\pm$ 0.026 & 0.16 $\pm$ $0.01$ \\
$\widetilde{f}_\|$ & & & & $0.435 \pm 0.045^{\dagger \dagger}$ & 0.34 $\pm$ $0.02$  \\
$\widetilde{f}_{\perp}$ & &  & & $0.401 \pm 0.040^{\dagger \dagger}$ & 0.500 $\pm$ $0.014$ \\
$f_0^{\rm av}$ & & 0.720 $\pm$ $0.029^{\dagger}$  & & & \\
$\mathcal{A}_{0}$ & & & & $-0.62 \pm 0.13^{\dagger \dagger}$ & \\
$\mathcal{A}_\|$ & & & & $0.188 \pm 0.043^{\dagger \dagger}$ & \\
$\mathcal{A}_{\perp}$ & & & & $0.050 \pm  0.042^{\dagger \dagger}$ & \\
$\mathcal{A}^{0}_{CP}$ & & 0.664 $\pm$ $0.088^{\dagger}$  & & & \\
$\mathcal{A}^\|_{CP}$ & & $-$0.063 $\pm$ $0.140^{\dagger}$  & & & \\
$\mathcal{A}^{\perp}_{CP}$ & & 0.284 $\pm$ $0.149^{\dagger}$  & &  &\\
$\tilde\delta_0$ & & & & $1.57\pm0.20^{\dagger\dagger}$& \\
$\tilde\delta_\|$ & & & &$0.795\pm0.074 ^{\dagger\dagger}$ & \\
$\tilde\delta_\perp$ & & & &$-2.365\pm0.063 ^{\dagger\dagger}$ & \\
$\delta_0^{CP}$ & & $0.360\pm 0.092^{\dagger}$ & & $0.12\pm0.09 ^{\dagger\dagger}$ & \\
$\delta_\|^{CP}$ & & $0.239 \pm 0.110^{\dagger}$ & & $0.014\pm0.040 ^{\dagger\dagger}$ & \\
$\delta_\perp^{CP}$ & & $0.206\pm0.109 ^{\dagger}$ & & $0.000\pm0.035 ^{\dagger\dagger}$ & \\
\hline
\end{tabular}}
\caption{Observables for $\Delta S=1$ $B \to VV$ decays in which $V \in \{\rho, K^* \}$. Data labeled $^\dagger$, $^{\dagger\dagger}$ and $^{\dagger\dagger\dagger}$ are taken from Refs.~\cite{LHCb:2025zvw}, \cite{LHCb:2018hsm}, and \cite{LHCb:2025ftm}, respectively. The remaining data are taken from Ref.~\cite{HFLAV:2024ctg}. The entry labeled $^*$ is the error-weighted average of the measurements from Refs.~\cite{LHCb:2025zvw} and \cite{BaBar:2010evf}. The tangent of the angles are used in the fits for better convergence.}
\label{tab:dataS1}
\end{table}

\begin{table} [H]
\centering
\setlength{\tabcolsep}{0.7pt}
\makebox[\textwidth]{
\begin{tabular}{|c|c|c|c|c|c|c|c|c|}
\hline
Decays & $B^+ \to \omega \rho^+$ & $B^+ \to \phi \rho^+$  & $B^0 \to \omega  \rho^0$ & $B^0 \to \phi \rho^0$ & $B^0 \to \omega \omega$ & $B^0 \to \omega \phi$ & $B^0 \to \phi \phi$ &  $\bs \to \phi \bar{K}^{*0}$  \\
\hline
${\cal B}_{CP}$ ($\times 10^{-6}$) & $15.9 \pm 2.1$ & $1.5 \pm 1.1^*$ & $0.8 \pm 0.5^\dagger$ & $0.09\pm0.16^*$  & $1.40\pm0.25^{\dagger\dagger}$& $0.0\pm0.3^{\dagger\dagger\dagger}$& $-0.04\pm0.12^*$&  $1.14\pm0.29$  \\
${\cal A}_{CP}$ & $-0.20\pm0.09$  & &  & & $-0.44\pm0.44^{\dagger\dagger\dagger\dagger}$ & & &  \\
$\widetilde{f}_{0}$ & $0.90\pm0.06$ & &  & & $0.87\pm0.18^{\dagger\dagger\dagger\dagger}$ & & & $0.51\pm0.17$ \\
$\widetilde{f}_{||}$ &  & &  & & & & & $0.21\pm0.11$\\
\hline
\end{tabular}}
\caption{
Observables for the $\Delta S=0$ $B \to VV$ decays in which at least one $V \in \{ \phi, \omega \}$. Data labeled $^*$,  $^{\dagger}$, $^{\dagger\dagger}$, $^{\dagger\dagger\dagger}$, and $^{\dagger\dagger\dagger\dagger}$ are taken from Refs.~\cite{BaBar:2008zea}, \cite{BaBar:2009mcf}, \cite{ParticleDataGroup:2026aaa}, \cite{BaBar:2013uwj}, and \cite{Belle:2024dhj}, respectively. The remaining data are taken from Ref.~\cite{HFLAV:2024ctg}.}
\label{tab:dataS0omega}
\end{table}

\begin{table} [H]
\centering
\setlength{\tabcolsep}{0.8pt}
\makebox[\textwidth]{
\begin{tabular}{|c|c|c|c|c|c|c|}
\hline
Observables & $B^+ \to \omega K^{*+}$ &  $B^+ \to \phi K^{*+}$ & $B^0 \to \omega K^{*0}$ & $B^0 \to \phi K^{*0}$ & $\bs \to \phi \rho^0$ & $\bs \to \phi \phi$   \\
\hline
${\cal B}_{CP}$ ($\times 10^{-6}$) & $2.4 \pm 1.0^*$ & 10.0 $\pm$ $1.1$ & 2.04 $\pm$ $0.49$ & 10.09 $\pm$ $0.49$ & $0.27 \pm 0.08$ &  $18.5 \pm 1.4$  \\
${\cal A}_{CP}$ & 0.29 $\pm$ $0.35$ & $-$0.01 $\pm$ $0.08$ & 0.45 $\pm$ $0.25$ & $-$0.001 $\pm$ $0.041$ & & \\
$\widetilde{f}_{0}$ & 0.41 $\pm$ $0.19$ & 0.50 $\pm$ $0.05$ & 0.69  $\pm$ $0.11$ & $0.497 \pm 0.017$  & & $0.378 \pm 0.013$  \\
$\widetilde{f}_\|$ & & & $0.22 \pm 0.21^\dagger$ & &  & \\
$\widetilde{f}_{\perp}$ & & $0.20 \pm 0.05$ & $0.10 \pm 0.13^\dagger$ & $0.224 \pm 0.015$ & & $0.293 \pm 0.010$ \\
$\mathcal{A}_{0}$ & & & $-0.13 \pm 0.30^\dagger$ & $0.003 \pm 0.038^{\dagger\dagger}$ & & \\
$\mathcal{A}_\|$ & & & $0.26 \pm 0.59^\dagger$ & & & \\
$\mathcal{A}_{\perp}$ & & & $0.3 \pm 0.9^\dagger$ & $-0.047 \pm 0.075^{\dagger\dagger}$ & & \\

$|\la_0|$ & & & & & & $1.02\pm0.17^{\dagger\dagger\dagger}$\\
$|\la_\|/\la_0|$ & & & & & & $0.78\pm0.21^{\dagger\dagger\dagger}$\\
$|\la_\perp/\la_0|$ & & & & & & $0.97\pm0.22^{\dagger\dagger\dagger}$\\
$\widetilde{\delta_{0}}$ & & & $-0.86 \pm 0.77^\dagger$ & & & \\
$\widetilde{\delta_{\|}}$ & & &$-1.83 \pm0.43 ^\dagger$ &  $2.562\pm0.080^{\dagger\dagger}$& &\\
$\widetilde{\delta_{\perp}}$ & & & $1.6 \pm 0.7 ^\dagger$ &$2.633\pm0.072^{\dagger\dagger}$ & & \\
${\delta_{0}^{CP}}$ & & & $ 0.03\pm 0.33^\dagger$ & & & \\
${\delta_{\|}^{CP}}$ & & & $0.59 \pm 0.30 ^\dagger$ & $-0.045\pm0.071^{\dagger\dagger}$& &\\
${\delta_{\perp}^{CP}}$& & & $-0.25 \pm 0.46^\dagger$ &$-0.062\pm0.062^{\dagger\dagger}$ && \\
$\delta_{\|}-\delta_0$& & & & & & $2.463\pm0.030^{\dagger\dagger\dagger}$\\
$\delta_{\perp}-\delta_0$& & & & & & $2.769\pm0.106^{\dagger\dagger\dagger}$\\
\hline
\end{tabular}}
\caption{
Observables for the $\Delta S=1$ $B \to VV$ decays in which at least one $V \in \{ \phi, \omega \}$. Data labeled $^*$, $^\dagger$, $^{\dagger\dagger}$, and $^{\dagger\dagger\dagger}$ are taken from Refs.~\cite{BaBar:2009mcf}, \cite{LHCb:2018hsm}, \cite{LHCb:2014xzf}, and \cite{LHCb:2023exl}, respectively. The remaining data are taken from Ref.~\cite{HFLAV:2024ctg}. The tangent of the angles are used in the fit for better convergence. Note that the angles given in \cite{LHCb:2014xzf} assume that $\delta_0 = 0$, which has been taken into account in the fit.} 
\label{tab:dataS1omega}
\end{table}

\newpage

\section{Fit results}
\label{app:fit_results}

Table \ref{tab:rhoKst} shows the best-fit values for the $B \to \rho K^*$ decays using isospin symmetry. These same decays are also studied with the additional constraint $|\widetilde{P_{uc}^{\lambda}}/\widetilde{P_{tc}^{\lambda}}| \le 1$ ($\lambda = 0,||,\perp$), and the results are shown in Table \ref{tab:rhoKstcons}. The $B \to VV$ decays, $V \in \{ \rho, K^* \}$, are studied using the SU(3)$_F$ symmetry and the fit results are shown in Table \ref{tab:SU3}. Table \ref{tab:SU3omega} shows the best-fit values for the $B \to VV$ decays, $V \in \{ \rho, K^*, \phi, \omega \}$, with the assumption of SU(3)$_F$ symmetry.

\begin{table}[H]
\centering
\renewcommand{\arraystretch}{0.8}
\setlength{\tabcolsep}{8pt}
\begin{tabular}{|c|c|c|c|c|c|}
    \hline
Parameter  & Best-fit value & Parameter  & Best-fit value & Parameter  & Best-fit value \\ 
\hline
$|\widetilde{T^0}|$ &  $45\pm8$  & $|\widetilde{T^\||}$ &  $32\pm11$  & $|\widetilde{T^{\perp}}|$ &  $23\pm8$  \\
  $|\widetilde{C^0}|$ &  $25\pm4$  & $|\widetilde{C^\|}|$ &  $14\pm6$  & $|\widetilde{C^{\perp}}|$ &  $24\pm4$ \\  
  $|\widetilde{P_{uc}^0}|$ &  $27\pm4$  & $|\widetilde{P_{uc}^\|}|$ &  $27.1\pm3.4$  & $|\widetilde{P_{uc}^{\perp}}|$ &  $9.6\pm1.5$  \\
  $|\widetilde{P_{tc}^0}|$ &  $0.118\pm0.035$  & $|\widetilde{P_{tc}^\|}|$ &  $0.165\pm0.023$  & $|\widetilde{P_{tc}^{\perp}}|$ &  $0.040\pm0.027$ 
 \\ $\delta_{\widetilde{C^0}}$ & $(157 \pm 14)^{\circ}$ & $\delta_{\widetilde{C^\|}}$ & $(332 \pm 15)^{\circ}$ & $\delta_{\widetilde{C^{\perp}}}$ & $(203 \pm 4)^{\circ}$ \\
  $\delta_{\widetilde{P_{uc}^0}}$ & $(143 \pm 14)^{\circ}$  & $\delta_{\widetilde{P_{uc}^\|}}$ &  $(256 \pm 13)^{\circ}$ & $\delta_{\widetilde{P_{uc}^{\perp}}}$ & $(109 \pm 23)^{\circ}$ \\
  $\delta_{\widetilde{P_{tc}^0}}$ & $(36 \pm 23)\times10^{\circ}$ & $\delta_{\widetilde{P_{tc}^\|}}$ & $(252 \pm 12)^{\circ}$& $\delta_{\widetilde{P_{tc}^{\perp}}}$ & $(139 \pm 32)^{\circ}$ \\
   $\delta_{\widetilde{T^0}}$ & $(299 \pm 14)^{\circ}$ &  $\delta_{\widetilde{T^\|}}$ & $(128 \pm 15)^{\circ}$ & & \\ 
    \hline
    \end{tabular} 
    \caption{Results of the fit to observables in $B \to \rho K^*$ decays. The magnitudes of diagrams for the three transversity components are allowed to vary freely, and the phases vary between 0 and $360^{\circ}$. We find $\chi^{2}_{\rm min}/{\rm d.o.f.} =$ \chirhoK/\dofrhoK, for a $p$-value of \pvalrhoK.}
    \label{tab:rhoKst}
\end{table}

\begin{table}[H]
\centering
\renewcommand{\arraystretch}{0.8}
\setlength{\tabcolsep}{9pt}
\begin{tabular}{|c|c|c|c|c|c|}
    \hline
Parameter  & Best-fit value & Parameter  & Best-fit value & Parameter  & Best-fit value \\ 
\hline
$|\widetilde{T^0}|$ &  $32.0\pm1.5$  & $|\widetilde{T^\||}$ &  $24.6\pm0.9$  & $|\widetilde{T^{\perp}}|$ &  $3.3\pm3.3$  \\
  $|\widetilde{C^0}|$ &  $6.4\pm0.9$  & $|\widetilde{C^\|}|$ &  $26.5\pm0.8$  & $|\widetilde{C^{\perp}}|$ &  $1.4\pm1.9$ \\  
  $|\widetilde{P_{uc}^0}|/|\widetilde{P_{tc}^0}|$ &  $1.0\pm0.6$  & $|\widetilde{P_{uc}^\|}|/|\widetilde{P_{tc}^\|}|$ &  $1.0\pm0.5$  & $|\widetilde{P_{uc}^{\perp}}|/|\widetilde{P_{tc}^\perp}|$ &  $1.0\pm0.9$  \\
  $|\widetilde{P_{tc}^0}|$ &  $0.227\pm0.018$  & $|\widetilde{P_{tc}^\|}|$ &  $0.029\pm0.016$  & $|\widetilde{P_{tc}^{\perp}}|$ &  $0.453\pm0.021$ 
 \\ $\delta_{\widetilde{C^0}}$ & $(347 \pm 9)^{\circ}$ & $\delta_{\widetilde{C^\|}}$ & $(36.8 \pm 1.2)^{\circ}$ & $\delta_{\widetilde{C^{\perp}}}$ & $(27 \pm 13)\times10^{\circ}$ \\
  $\delta_{\widetilde{P_{uc}^0}}$ & $(6 \pm 20)\times10^{\circ}$  & $\delta_{\widetilde{P_{uc}^\|}}$ &  $(36 \pm 29)\times10^{\circ}$ & $\delta_{\widetilde{P_{uc}^{\perp}}}$ & $(32 \pm 28)\times10^{\circ}$ \\
  $\delta_{\widetilde{P_{tc}^0}}$ & $(63 \pm 7)^{\circ}$ & $\delta_{\widetilde{P_{tc}^\|}}$ & $(36 \pm 35)\times10^{\circ}$& $\delta_{\widetilde{P_{tc}^{\perp}}}$ & $(3 \pm 7)\times10^{\circ}$ \\
   $\delta_{\widetilde{T^0}}$ & $(297 \pm 10)^{\circ}$ &  $\delta_{\widetilde{T^\|}}$ & $(216.7 \pm 1.2)^{\circ}$ & & \\ 
    \hline
    \end{tabular} 
    \caption{Fit results for the $B \to \rho K^*$ decays in which we have fixed $|\widetilde{P_{uc}^{\lambda}}/\widetilde{P_{tc}^{\lambda}}| \le 1$ ($\lambda = 0,||,\perp$). The phases vary between 0 and $360^{\circ}$. We find $\chi^{2}_{\rm min}/{\rm d.o.f.} =$ \chirhoKConstraint/\dofrhoK, for a $p$-value of \pvalrhoKConstraint.}
    \label{tab:rhoKstcons}
\end{table}

\newpage

\begin{table}[htb!]
\centering
\renewcommand{\arraystretch}{0.8}
\setlength{\tabcolsep}{10pt}
\makebox[\textwidth]{
\begin{tabular}{|c|c|c|c|c|c|}
    \hline
Parameter  & Fit result & Parameter  & Fit result & Parameter  & Fit result \\ 
\hline
  $|\widetilde{T^0}|$ & $16.9\pm1.0$ & $|\widetilde{T^\||}$ & $2.1\pm1.4$ & $|\widetilde{T^{\perp}}|$ & $2.7\pm1.4$ \\
  $|\widetilde{C^0}|$ & $3.9\pm1.0$ & $|\widetilde{C^\|}|$ & $1.0\pm1.3$ & $|\widetilde{C^{\perp}}|$ & $2.0\pm1.2$ \\  
  $|\widetilde{P_{uc}^0}|$ & $1.0\pm1.1$ & $|\widetilde{P_{uc}^\|}|$ &  $1.5\pm1.4$ & $|\widetilde{P_{uc}^{\perp}}|$ & $1.7\pm1.3$ \\
  $|\widetilde{A^0}|$ & $3.4\pm1.3$ & $|\widetilde{A^\|}|$ & $0.7\pm1.2$ & $|\widetilde{A^{\perp}}|$ & $1.3\pm1.4$ \\  
  $|\widetilde{PA_{uc}^0}|$ & $2.3\pm1.0$ & $|\widetilde{PA_{uc}^\|}|$ & $0.7\pm1.1$ & $|\widetilde{PA_{uc}^{\perp}}|$ & $0.9\pm0.9$ \\
  $|\widetilde{P_{tc}^0}|$ & $0.451\pm0.028$ & $|\widetilde{P_{tc}^\|}|$ &  $0.333\pm0.027$ & $|\widetilde{P_{tc}^{\perp}}|$ & $0.327\pm0.027$ \\
  $|\widetilde{PA_{tc}^0}|$ & $0.20\pm0.14$ & $|\widetilde{PA_{tc}^\|}|$ & $0.5\pm0.4$ & $|\widetilde{PA_{tc}^{\perp}}|$ & $0.62\pm0.32$ \\ 
  $\delta_{\widetilde{C^0}}$ & $(70 \pm 24)^{\circ}$ & $\delta_{\widetilde{C^\|}}$ & $(5 \pm 8)\times10^{\circ}$ & $\delta_{\widetilde{C^{\perp}}}$ & $(8 \pm 4)\times10^{\circ}$ \\
  $\delta_{\widetilde{P_{uc}^0}}$ & $(23 \pm 8)\times10^{\circ}$  & $\delta_{\widetilde{P_{uc}^\|}}$ &  $(2 \pm 6)\times10^{\circ}$ & $\delta_{\widetilde{P_{uc}^{\perp}}}$ & $(10 \pm 5)\times10^{\circ}$ \\
  $\delta_{\widetilde{A^0}}$ & $(28 \pm 33)^{\circ}$ & $\delta_{\widetilde{A^\|}}$ & $(18 \pm 14)\times10^{\circ}$ & $\delta_{\widetilde{A^{\perp}}}$ & $(26 \pm 7)\times10^{\circ}$\\ 
  $\delta_{\widetilde{PA_{uc}^0}}$ & $(20 \pm 28)^{\circ}$  & $\delta_{\widetilde{PA_{uc}^\|}}$ &  $(19 \pm 10)\times10^{\circ}$ & $\delta_{\widetilde{PA_{uc}^{\perp}}}$ & $(26 \pm 7)\times10^{\circ}$ \\
  $\delta_{\widetilde{P_{tc}^0}}$ & $(164 \pm 18)^{\circ}$ & $\delta_{\widetilde{P_{tc}^\|}}$ & $(45 \pm 5)^{\circ}$& $\delta_{\widetilde{P_{tc}^{\perp}}}$ & $(130 \pm 4)^{\circ}$ \\
  $\delta_{\widetilde{PA_{tc}^0}}$ & $(1 \pm 5)\times10^{\circ}$ & $\delta_{\widetilde{PA_{tc}^\|}}$ & $(16 \pm 6)\times10^{\circ}$& $\delta_{\widetilde{PA_{tc}^{\perp}}}$ & $(24 \pm 5)\times10^{\circ}$ \\
  $\delta_{\widetilde{T^0}}$ & $(142 \pm 20)^{\circ}$ &  $\delta_{\widetilde{T^\|}}$ & $(22 \pm 4)\times10^{\circ}$ & & \\   
    \hline
    \end{tabular}  }
    \caption{Fit results for the $B\to VV$ decays, with $V \in \{\rho, K^*\}$. The magnitudes of diagrams for the three transversity components are allowed to vary freely, and the phases vary between 0 and $360^{\circ}$. We find $\chi^{2}_{\rm min}/{\rm d.o.f.} =$ \chiVV/\dofVV, for a $p$-value of \pvalVV.
    }
    \label{tab:SU3}
\end{table}

\newpage

\begin{table}[htb!]
\centering
\renewcommand{\arraystretch}{0.8}
\setlength{\tabcolsep}{8pt}
\makebox[\textwidth]{
\begin{tabular}{|c|c|c|c|c|c|c|}
    \hline
Type & Parameter  & Fit result & Parameter  & Fit result & Parameter  & Fit result \\ 
\hline
\multirow{14}{*}{\rotatebox{90}{$(\boldsymbol{8\otimes 8})_S$}} 
&$|\widetilde{T^0}|$ & $16.3\pm0.9$ & $|\widetilde{T^\||}$ & $2.9\pm0.9$ & $|\widetilde{T^{\perp}}|$ & $3.0\pm0.9$ \\
&  $|\widetilde{C^0}|$ & $4.2\pm0.7$ & $|\widetilde{C^\|}|$ & $0.7\pm0.9$ & $|\widetilde{C^{\perp}}|$ & $1.8\pm0.7$ \\  
&  $|\widetilde{P_{uc}^0}|$ & $1.4\pm0.7$ & $|\widetilde{P_{uc}^\|}|$ &  $1.2\pm0.8$ & $|\widetilde{P_{uc}^{\perp}}|$ & $1.2\pm0.7$ \\
&  $|\widetilde{A^0}|$ & $3.8\pm0.8$ & $|\widetilde{A^\|}|$ & $0.7\pm1.1$ & $|\widetilde{A^{\perp}}|$ & $0.4\pm0.7$ \\  
&  $|\widetilde{PA_{uc}^0}|$ & $2.6\pm0.6$ & $|\widetilde{PA_{uc}^\|}|$ & $0.3\pm0.9$ & $|\widetilde{PA_{uc}^{\perp}}|$ & $0.2\pm0.6$ \\
&  $|\widetilde{P_{tc}^0}|$ & $0.563\pm0.022$ & $|\widetilde{P_{tc}^\|}|$ &  $0.418\pm0.023$ & $|\widetilde{P_{tc}^{\perp}}|$ & $0.406\pm0.024$ \\
&  $|\widetilde{PA_{tc}^0}|$ & $0.28\pm0.05$ & $|\widetilde{PA_{tc}^\|}|$ & $0.36\pm0.13$ & $|\widetilde{PA_{tc}^{\perp}}|$ & $0.61\pm0.09$ \\ 
&  $\delta_{\widetilde{C^0}}$ & $(87 \pm 17)^{\circ}$ & $\delta_{\widetilde{C^\|}}$ & $(13 \pm 9)\times10^{\circ}$ & $\delta_{\widetilde{C^{\perp}}}$ & $(95 \pm 26)^{\circ}$ \\
&  $\delta_{\widetilde{P_{uc}^0}}$ & $(22 \pm 4)\times10^{\circ}$  & $\delta_{\widetilde{P_{uc}^\|}}$ &  $(9 \pm 5)\times10^{\circ}$ & $\delta_{\widetilde{P_{uc}^{\perp}}}$ & $(12 \pm 4)\times10^{\circ}$ \\
&  $\delta_{\widetilde{A^0}}$ & $(46 \pm 23)^{\circ}$ & $\delta_{\widetilde{A^\|}}$ & $(23 \pm 9)\times10^{\circ}$ & $\delta_{\widetilde{A^{\perp}}}$ & $(28 \pm 27)\times10^{\circ}$\\ 
&  $\delta_{\widetilde{PA_{uc}^0}}$ & $(36 \pm 18)^{\circ}$  & $\delta_{\widetilde{PA_{uc}^\|}}$ &  $(24 \pm 27)\times10^{\circ}$ & $\delta_{\widetilde{PA_{uc}^{\perp}}}$ & $(21 \pm 14)\times10^{\circ}$ \\
&  $\delta_{\widetilde{P_{tc}^0}}$ & $(175 \pm 9)^{\circ}$ & $\delta_{\widetilde{P_{tc}^\|}}$ & $(143 \pm 8)^{\circ}$& $\delta_{\widetilde{P_{tc}^{\perp}}}$ & $(147 \pm 7)^{\circ}$ \\
&  $\delta_{\widetilde{PA_{tc}^0}}$ & $(12 \pm 25)^{\circ}$ & $\delta_{\widetilde{PA_{tc}^\|}}$ & $(248 \pm 13)^{\circ}$& $\delta_{\widetilde{PA_{tc}^{\perp}}}$ & $(264 \pm 9)^{\circ}$ \\
&  $\delta_{\widetilde{T^0}}$ & $(150 \pm 11)^{\circ}$ &  $\delta_{\widetilde{T^\|}}$ & $(311 \pm 22)^{\circ}$ & & \\
    \hline \hline
\multirow{8}{*}   {\rotatebox{90}{$(\boldsymbol{8\otimes 1})$}}  &$|\widetilde{T^0}|$ & $8.3\pm2.4$ & $|\widetilde{T^\||}$ & $6\pm6$ & $|\widetilde{T^{\perp}}|$ & $4\pm5$ \\
&  $|\widetilde{C^0}|$ & $4.2\pm1.8$ & $|\widetilde{C^\|}|$ & $0\pm6$ & $|\widetilde{C^{\perp}}|$ & $0.6\pm1.4$ \\  
&  $|\widetilde{P_{uc}^0}|$ & $2.7\pm1.0$ & $|\widetilde{P_{uc}^\|}|$ &  $0.7\pm1.1$ & $|\widetilde{P_{uc}^{\perp}}|$ & $0.7\pm1.2$ \\
&  $|\widetilde{P_{tc}^0}|$ & $0.568\pm0.022$ & $|\widetilde{P_{tc}^\|}|$ &  $0.362\pm0.028$ & $|\widetilde{P_{tc}^{\perp}}|$ & $0.304\pm0.026$ \\
&  $\delta_{\widetilde{C^0}}$ & $(49 \pm 26)^{\circ}$ & $\delta_{\widetilde{C^\|}}$ & $(25 \pm 20)\times10^{\circ}$ & $\delta_{\widetilde{C^{\perp}}}$ & $(9 \pm 21)\times10^{\circ}$ \\
&  $\delta_{\widetilde{P_{uc}^0}}$ & $(175 \pm 29)^{\circ}$  & $\delta_{\widetilde{P_{uc}^\|}}$ &  $(7 \pm 21)\times10^{\circ}$ & $\delta_{\widetilde{P_{uc}^{\perp}}}$ & $(23 \pm 27)\times10^{\circ}$ \\
&  $\delta_{\widetilde{P_{tc}^0}}$ & $(184 \pm 10)^{\circ}$ & $\delta_{\widetilde{P_{tc}^\|}}$ & $(150 \pm 5)^{\circ}$& $\delta_{\widetilde{P_{tc}^{\perp}}}$ & $(155 \pm 6)^{\circ}$ \\
&  $\delta_{\widetilde{T^0}}$ & $(12 \pm 5)\times10^{\circ}$ &  $\delta_{\widetilde{T^\|}}$ & $(31 \pm 10)\times10^{\circ}$ &$\delta_{\widetilde{T^\perp}}$ &$(20 \pm 27)\times10^{\circ}$ \\
\hline \hline
\multirow{4}{*}   {\rotatebox{90}{$(\boldsymbol{1\otimes 1})$}} &  $|\widetilde{C^0}|$ & $3.7\pm2.2$ & $|\widetilde{C^\|}|$ & $0.4\pm2.9$ & $|\widetilde{C^{\perp}}|$ & $0.5\pm2.6$ \\  
&  $|\widetilde{P_{tc}^0}|$ & $0.7\pm0.6$ & $|\widetilde{P_{tc}^\|}|$ &  $0.8\pm0.6$ & $|\widetilde{P_{tc}^{\perp}}|$ & $1.1\pm0.5$ \\
&  $\delta_{\widetilde{C^0}}$ & $(12 \pm 5)\times10^{\circ}$ & $\delta_{\widetilde{C^\|}}$ & $(0 \pm 4)\times100^{\circ}$ & $\delta_{\widetilde{C^{\perp}}}$ & $(29 \pm 33)\times10^{\circ}$ \\
&  $\delta_{\widetilde{P_{tc}^0}}$ & $(253 \pm 30)^{\circ}$ & $\delta_{\widetilde{P_{tc}^\|}}$ & $(96 \pm 28)^{\circ}$& $\delta_{\widetilde{P_{tc}^{\perp}}}$ & $(134 \pm 27)^{\circ}$ \\
\hline
\end{tabular} }
    \caption{Fit results for the $B\to VV$ decays, with $V \in \{\rho, K^*,\phi, \omega \}$. The magnitudes of diagrams for the three transversity components are allowed to vary freely, and the phases vary between 0 and $360^\circ$. We find $\chi^{2}_{\rm min}/{\rm d.o.f.} =$ \chiAll/\dofAll, which corresponds to a $p$-value \pvalAll.
    }
    \label{tab:SU3omega}
\end{table}

\section{\boldmath Significant contributions to the $\chi^2$}
\label{app:contrib_chi2}

In this Appendix, we present the observables that contribute the most to the $\chi^2$ for the three poor fits. To be specific, we focus on the fits of $B\to \rho K^*$ with the constraint $|\widetilde{P_{uc}^{\lambda}}/\widetilde{P_{tc}^{\lambda}}| \le 1$, $B\to VV$ with $V \in \{\rho, K^*\}$, and $B\to VV$ with $V \in \{\rho, K^*, \omega, \phi\}$.

\begin{figure}[H]
    \centering
    \includegraphics[width=\linewidth]{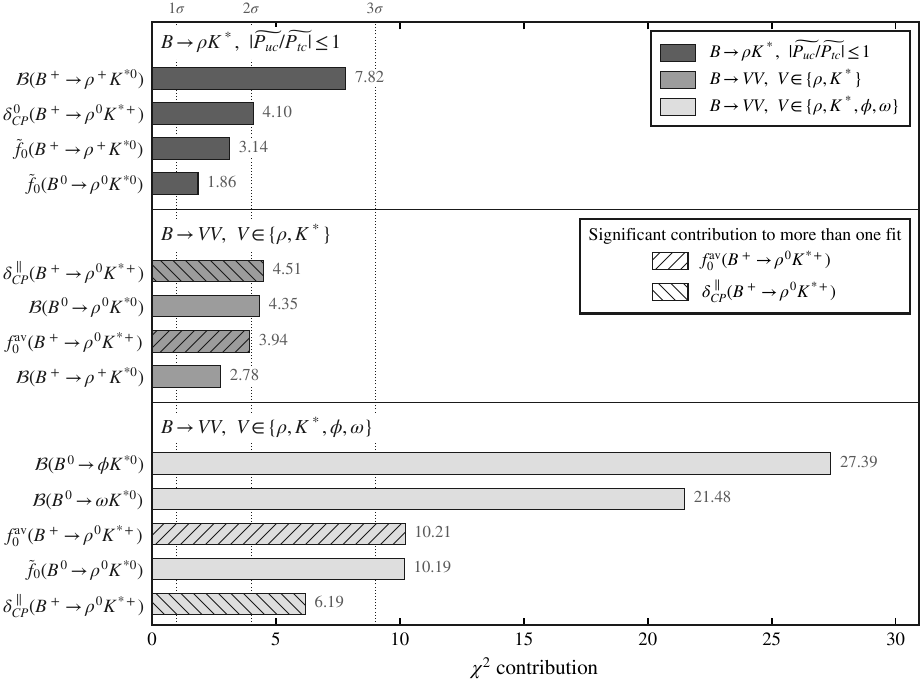}
    \caption{For each of the three poor fits, the figure shows the four observables making the largest contributions to the $\chi^2$ when evaluated at the best-fit point (see Appendix \ref{app:fit_results}). We show five observables for the full SU(3)$_F$ fit since the contribution of the fifth largest observable is larger than all the contributions to the $B\to VV$ fit with $V\in\{\rho, K^*\}$. 
    For each observable, the shading identifies the fit(s) to which it contributes, ranging from the darkest shade for the constrained $B\to\rho K^*$ fit to the lightest for the $B \to VV$ with $\omega$ and $\phi$. Diagonal hatching identifies observables that make large contributions to more than one fit.}
    \label{fig:chi2contribution}
\end{figure}

\newpage

In Fig.~\ref{fig:chi2contribution}, we present the observables that make the largest contributors to the $\chi^2$ for the three fits. In general, these observables make large contributions to only one fit. For example, in the $B\to VV$ fit with $\phi$ and $\omega$, the main contributors are observables unique to this fit. Still, there are two exceptions. $f_0^{\text{av}}(B^+\to \rho^0K^{*+})$ appears in the $B\to VV$ fits with $V\in\{\rho, K^*\}$ and $V\in\{\rho, K^*, \phi, \omega\}$ (though its contribution to the second fit is more than twice as large as that in the first fit). And $\delta_{CP}^\|(B^+\to \rho^0K^{*+})$ also contributes to both SU(3)$_F$ fits. In fact, the largest contributors fitted in both SU(3)$_F$ fits are all $B\to\rho K^*$ observables.

\section{Relations between observables}\label{app:relations}

The amplitude for any $B\to VV$ decay is expressed in terms of three polarization amplitudes. Furthermore, each polarization amplitude can, in principle, have two parts, proportional to two different combinations of CKM matrix elements. Therefore, the amplitude for a general $B\to VV$ decay depends on six complex hadronic amplitudes, or 11 real parameters -- six magnitudes and five relative phases. However, in Ref.~\cite{Bhattacharya:2013sga}, it was shown that a time-dependent angular analysis can be used in $B\to VV$ decays to isolate as many as 24 observable angular coefficients. It was also shown there that not all observable coefficients are independent. Six relationships were identified. Still, there are many more relationships between the observables measured in $B\to VV$ decays. Here we will identify several additional relationships. 

Let us consider the polarization fractions given in Eqs. (\ref{eq:polf}) and (\ref{polfracCPasym}), and repeated here for convenience.
\beq
f_\la ~=~ \frac{|A_\la|^2}{\sum\limits_\la|A_\la|^2} ~,~~
{\bar f}_\la ~=~ \frac{|{\bar A}_\la|^2}{\sum\limits_\la|{\bar A}_\la|^2} ~,~~ {\tilde f}_\la ~=~ \frac12(f_\la + {\bar f}_\la)~.~~
\eeq
These polarization fractions all satisfy the following relationship.
\beq
\sum\limits_\la f_\la ~=~ 
\sum\limits_\la {\bar f}_\la ~=~ 
\sum\limits_\la {\tilde f}_\la ~=~ 1~.~~ 
\eeq
Therefore, even though several combinations of the polarization fractions can be measured, there are only four independent observables among such measurements. 

The $CP$-averaged branching fraction, ${\cal B}_{\rm CP}$, defined in Eq.~(\ref{eq:bcp}), is another independent observable, and so is the direct $CP$-asymmetry, ${\cal A}_{\rm CP}$, defined in Eq.~(\ref{directCPasym}). However, several additional observables used in our fits are not independent. Similar to the polarization fractions, one can also measure the following polarization-dependent direct $CP$ asymmetries [see Eqs.~(\ref{directCPasym}) and (\ref{polfracCPasym})] and $CP$-averaged polarization fractions [see Eq.~(\ref{eq:polfCPavg})].
\beq
{\cal A}_{CP}^\la ~=~ \frac{|{\bar A}_\la|^2 - |A_\la|^2}{|{\bar A}_\la|^2 + |A_\la|^2}~,~~~ {\cal A}_\la ~=~ \frac{{\bar f}_\la - f_\la}{{\bar f}_\la + f_\la}~,~~~ f_\la^{av} ~=~ \frac{|A_\la|^2 + |{\bar A}_\la|^2}{\sum\limits_\la(|A_\la|^2 + |{\bar A}_\la|^2)}~.~~
\eeq
Not all of these observables are independent. First, the three ${\cal A}_\la$'s and the three ${\tilde f}_\la$'s are related as follows:
\beq
\sum\limits_\la {\cal A}_\la\,{\tilde f}_\la ~=~ \frac12\sum\limits_\la({\bar f}_\la - f_\la) ~=~ 0\,.
\eeq
Second, ${\cal A} _{CP}$ can be expressed as a sum of products of the three $f_\la^{av}$'s and the three ${\cal A}^\la_{CP}$'s as follows:
\bea
{\cal A}_{CP} = \frac{|{\bar A}|^2 - |A|^2}{|{\bar A}|^2 + |A|^2} = \sum_\la\frac{|{\bar A}_\la|^2 - |A_\la|^2}{\sum\limits_\la(|{\bar A}_\la|^2 + |A_\la|^2)} = \sum_\la{\cal A}^\la_{CP}\,\frac{|{\bar A}_\la|^2 + |A_\la|^2}{\sum\limits_\la(|{\bar A}_\la|^2 + |A_\la|^2)} = \sum_\la{\cal A}^\la_{CP}\,f^{av}_\la.~~~~~~
\eea
Note that each ${\cal A}^i_{CP}$ (here $i$ represents polarization) is related to the magnitude $|\la_i| = |{\bar A}_i/A_i|$ defined in Eq.~(\ref{eq:ladef}), as follows
\beq
{\cal A}^i_{CP} = \frac{|\la_i|^2-1}{|\la_i|^2+1}~.
\eeq

Finally, a polarization-dependent relationship exists between $f_\la^{av}$ and ${\tilde f}_\la$. Since in the absence of $CP$ violation these two quantities are expected to be equal to each other, we expect their difference to be proportional to a $CP$-violating observable. Indeed, we find that the difference is proportional to ${\cal A}_{CP}$:
\bea
f_\la^{av} - {\tilde f}^{}_\la &=& \frac{|A_\la|^2 + |{\bar A}_\la|^2}{|A|^2 + |{\bar A}|^2} - \frac12 \left(\frac{|A_\la|^2}{|A|^2} + \frac{|{\bar A}_\la|^2}{|{\bar A}|^2}\right) ~,~~ \nl 
&=& \frac12\frac{|{\bar A}|^2 - |A|^2}{|{\bar A}|^2 + |A|^2}\left(\frac{{|\bar A}_\la|^2}{|{\bar A}|^2} - \frac{|A_\la|^2}{|{ A}|^2}\right) ~=~ \,{\cal A}_{CP}\,\mathcal{A}_\lambda {\tilde f}^{}_\la ~.~~
\eea

Depending on the decay, a variety of different combinations of observables are available from experimental studies. In our fits, we count the number of independent observables in each decay to obtain the corresponding number of degrees of freedom. We found that the number of independent observables available for our fits in the $B^0\to K^{*0}\bar K^{*0}$, $B^+\to \rho^0K^{*+}$, $B^0\to \rho^0K^{*0}$, $B^0_s\to K^{*0}\bar K^{*0}$, and $B^0\to\omega K^{*0}$ channels were, respectively, three, nine, eleven, three, and eleven. In all other decays, the measured observables available for our fits were all independent.

\bibliography{b2rhoKst}
\bibliographystyle{apsrev4-2}

\end{document}